\documentclass[a4paper,11pt]{article}
\pdfoutput=1

\usepackage{jinstpub}
\usepackage{xcolor}
\usepackage{floatrow}

\usepackage{subfig}

\title{\boldmath Quality Assurance Test of Silicon Photomultipliers  and Electronic Boards for STAR Event Plane Detector}
\author[a,b]{Ming Shao,}
\author[a,b]{Yitao Wu,}
\author[a,b]{Zheng Liang,}
\author[a,b]{Kaifeng Shen,}
\author[a,b]{Zebo Tang,}
\author[c]{M.A. Lisa,}
\author[d]{R. Reed,}
\author[e]{G. Visser,}
\author[a,b]{Yongjie Sun,}
\author[a,b]{Yi Zhou,}
\author[a,b]{Jian Zhou,}
\author[a,b]{Guofeng Song,}
\author[a,b]{Dongdong Hu,}
\author[a,b]{Xu Wang,}
\author[a,b]{Xinjian Wang}
\affiliation[a]{State Key Laboratory of Particle Detection and Electronics, University of Science and Technology of China, Hefei 230026, China}
\affiliation[b]{Department of Modern Physics, University of Science and Technology of China, Hefei 230026, China}
\affiliation[c]{The Ohio State University, Columbus, Ohio 43210, USA}
\affiliation[d]{Lehigh University, Bethlehem, Pennsylvania, 18015, USA}
\affiliation[e]{Indiana University, Bloomington, Indiana 47408, USA}
\emailAdd{lz1021@mail.ustc.edu.cn}

\abstract{
The event plane detector (EPD), installed in the Solenoid Tracker at the Relativistic Heavy-Ion Collider located at the Brookhaven National Laboratory is a plastic scintillator-based device that measures the reaction centrality and event plane in the forward region of the relativistic heavy-ion collisions.
We used silicon photomultiplier (SiPM) arrays to detect the photons produced in the scintillator via the fiber connection.
Signals from the SiPM arrays were amplified by the front-end electronic (FEE) board, and sent to {\color{black} the analog-to-digital converter (ADC) boards} for further processing via the receiver (RX) board.
The full EPD system consisted of 24 {\color{black} super-sectors (SSs)}; each SS was equipped with two SiPM boards, two FEE boards and two RX boards, and they corresponded to 744 readout channels.
All these boards {\color{black} were} mass produced at the University of Science and Technology of China, with a dedicated quality assurance (QA) procedures applied to {\color{black} identify any problems before deployment}.
This article describes the details of the QA method and the related test system.
The QA test results are presented along with the discussions.
}


\begin{document}
\maketitle
\flushbottom

\section{Introduction}
\label{sec:Introduction}


One of the main physics aims of the Relativistic Heavy-Ion Collider (RHIC) at Brookhaven National Lab (BNL) is to discover and study the {\color{black} properties} of quark-gluon plasma (QGP), which is a state of strongly acting matter that exists only under extreme pressure and energy density~\cite{HRNNC,EATC}.
The phase structure, especially the phase boundary and the anticipated critical end-point (CEP), of the nuclear matter {\color{black} is of particular interest} to researchers.
During the beam energy scan (BES) program at RHIC there have been hints of a CEP and first-order phase transition at center-of-mass energies ranging from 5 GeV to 30 GeV~\cite{RFH}.
However, many key measurements for locating the CEP and determining the phase transition were limited by the event plane resolution, statistics, and centrality determination by the time projection chamber (TPC) only, which may have caused unwanted auto-correlation in the analyses. 
Therefore, in phase II of the BES program {\color{black} (BES-II),} it {\color{black} was} proposed to take data with increased statistics and upgraded detectors in the forward region, which would help to further investigate the nuclear phase structure.
{\color{black} To explore the rich science of BES-II, the Soleniod Tracker at RHIC (STAR) was upgraded with the event plane detector (EPD), which is a highly segmented device with broad coverage at forward rapidity.
{
    \color{black} The EPD allows for the precise determination of event planes through measuring the hit positions of minimum-ionizing particles (MIP), and provides a measure of collision centrality in a kinematic range that avoids autocorrelations ~\cite{cyang}.
}

{\color{black} The EPD {\color{black}~\cite{EPD_NIM}} is made up of plastic scintillator with wavelength shifting fibers and SiPM for readout.} It consists of two wheels that extend from a radius of 4.6 cm to 91 cm and are placed at $z = \pm 375$ cm {\color{black} relative} to the center of STAR, which corresponds to $5.1 < |\eta| < 2.1$.
{\color{black} Each wheel is made up of 12 wedge-shaped super-sectors (SS) that consists of 2 rows of 15 tiles and a single innermost tile, which gives a total of 31 tiles per SS, as depicted in Figure~\ref{fig:EPD-Wheel}.}
Thus there are 372 channels on each side and a total of 744 channels.
The segmentation, size, and shape of the EPD design maximized the event plane resolution, centrality estimation and flow harmonic measurements {\color{black}~\cite{EPD_NIM, EPD_proposal}}.
Embedded in the scintillator are the wavelength shifting (WLS) fibers that guide the light from each tile up to a connector at the outer edge of the SS.
These WLS fibers were then coupled with two bundles of clear optical fibers which were finally {\color{black} air-coupled} to arrays of silicon photomultipliers (SiPMs) that converted and amplified the photons generated by the charged particle crossing the EPD.

\begin{figure}
    \centering
    \includegraphics[width=0.7\textwidth]{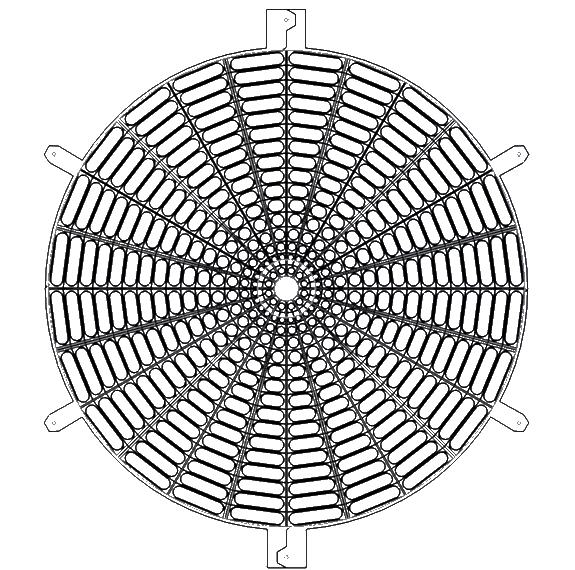}
    \caption{EPD layout.}
    \label{fig:EPD-Wheel}
\end{figure}

The construction of EPD {\color{black} began} in 2016.
A quarter of one EPD disc was installed in STAR and commissioned in 2017.
After verifying the main performance characteristics, the full production of EPD was launched in {\color{black} early} 2017 and completed by {\color{black} the end of that year}.
During this campaign, the heavy-ion group of the University of Science and Technology of China (USTC) had taken on the task of producing all the front-end boards, including the SiPM boards, the front-end electronics (FEE) boards and the receiver (RX) boards.
To ensure the high-quality functionality and safe operation of these boards, we had developed a dedicated test stand and quality assurance (QA) procedure, which consisted of a test platform, a graphical user interface (GUI) controller and a web-based database.
In section~\ref{sec:Boards}, we will give a brief introduction to these electronic boards.
And in sections~\ref{sec:ElectronicsQA} and~\ref{sec:BatchTest}, we will present the details of the QA process and the test results of the EPD mass-production.

\section{Electronics Boards}
\label{sec:Boards}

For each charged particle passing through the sensitive layer (plastic scintillator) of the EPD, the scintillation light produced in the scintillator will be detected by the SiPM arrays via a fiber connection.
The output signals from the SiPMs are then processed by the front-end electronic (FEE) board, which is closely connected to the SiPM board.
After pulse shaping, the signals are sent to {\color{black} the analog-to-digital converter (ADC) boards} for further processing and analysis via the receiver (RX) board that matches the electrical properties at the front and rear ends.

Hamamatsu MPPC (SiPM) S13360-1325PE was chosen as the photon sensor because of its high quantum efficiency (QE) in the blue-green light wavelength range, low noise rate, good linearity, intrinsic fast rise time 
and modest cost.
The sensitive area 1.3 mm $\times$ 1.3 mm fitted the size of clear fiber (1.1 mm in diameter) quite well.
The relatively small 25 $\rm \mu$m pixel pitch ensured a linearity of up to approximately 1000 photons, corresponding to approximately 10 minimum-ionizing particles (MIPs) hitting one EPD tile that is considered the upper limit of the normal EPD operation~\cite{EPD_proposal}.

The fiber-SiPM optical coupling was achieved by a properly designed connector, spacer and SiPM assembly board, as illustrated in Figure~\ref{fig:SiPM Coupling}.
Each SiPM board is 2 cm $\times$ 5 cm in size, populated with 16 SiPMs (see also Figure~\ref{fig:SiPM Photo}), and connected to a 16-fiber bundle. 
The spacer block is between the flat-faced fiber bundle connector and the SiPM board, which relieved the SiPMs from the stress of joining.
No coupling silicone oil or grease was used, and the fibers were air-coupled to the SiPMs to ensure long-term stability.
A thermal sensor was installed to the SiPM board to monitor the local temperature of the SiPMs, which {\color{black} was used to compensate the bias voltage of the SiPM. }


The SiPM board was connected to the FEE board using gold fingers, as shown in Figure~\ref{fig:SiPM Photo}.
In the lower-left corner of the plot a SiPM board was directly plugged into the housing connector of the FEE board.
The FEE board provides SiPM with {\color{black} an adjustable bias voltage (typically approximately 56 V) } and reads back the output signal from SiPM, and the local temperature.
{\color{black} 
The SiPM signal was amplified by a "regulated common base" amplifier with low input impedance, which maintained a more constant voltage on the SiPM for more constant gain and gathered the charge out of the SiPM more quickly.
After the first stage amplifier the signal was shaped by an L-C shaper with a peaking time of 16 ns, and then further amplified to drive the output.
The bias voltage can be set individually  for each SiPM, with compensations from the temperature feedback obtained from a single  temperature sensor (negative temperature coefficient (NTC) thermistor) on the SiPM board with an adjustable compensation slope.
The individual SiPM currents can be monitored with a 10 nA resolution and a range of 41 uA.
{
    \color{black}
    Meanwhile, the FEE board is capable of adjusting the baseline offsets of the output signals within a certain range, which is required by the ADC board used in STAR.
    By setting a specific digital-analog converter (DAC) on the board to values between -128 and 127, the FEE circuit can tune the current level to 0, which means a pedestal voltages 0 $\rm mV$ on 50 $\rm \Omega$ load at the RX board output.
}
The SiPM board temperature can also be monitored.
Via on-board digital-to-analog converters (DACs) and ADCs, the {\color{black}bias voltage} and {\color{black}dark current} are set and monitored remotely for each channel.

Control communications were performed by the I$^2$C \footnote{A multi-master, multi-slave, single-ended, serial computer bus invented by Philips Semiconductor (now NXP Semiconductors)} bus, with 12 FEE boards on a single bus in the installed system at STAR.
For the test stand, a single FEE  was controlled by the I$^2$C from the test computer and connected by means of a LinkUSB \footnote{Copyright by iButtonLink LLC., East Troy, Wi USA} interface under the 1-Wire\footnote{A trademark of Dallas Semiconductor Corp, Dallas, Texas, USA.} bus protocol. 
The ouput signals from  the FEE board were sent in differential mode to the RX board, which converted the differential signal to unipolar signal, required to interface with the standard STAR ADC system. 
Differential signals were used on the long cables between the FEE and RX boards because the cables and connectors were compact, and helped avoid difficulties with the interference and ground loops. 
The RX board outputs were connected to the ADC boards by short coaxial cables.
}

\begin{figure}
    \centering
    \begin{floatrow}
        \ffigbox{\caption{Schematic of SiPM and fiber coupling}\label{fig:SiPM Coupling}}{\includegraphics[width=0.45\textwidth]{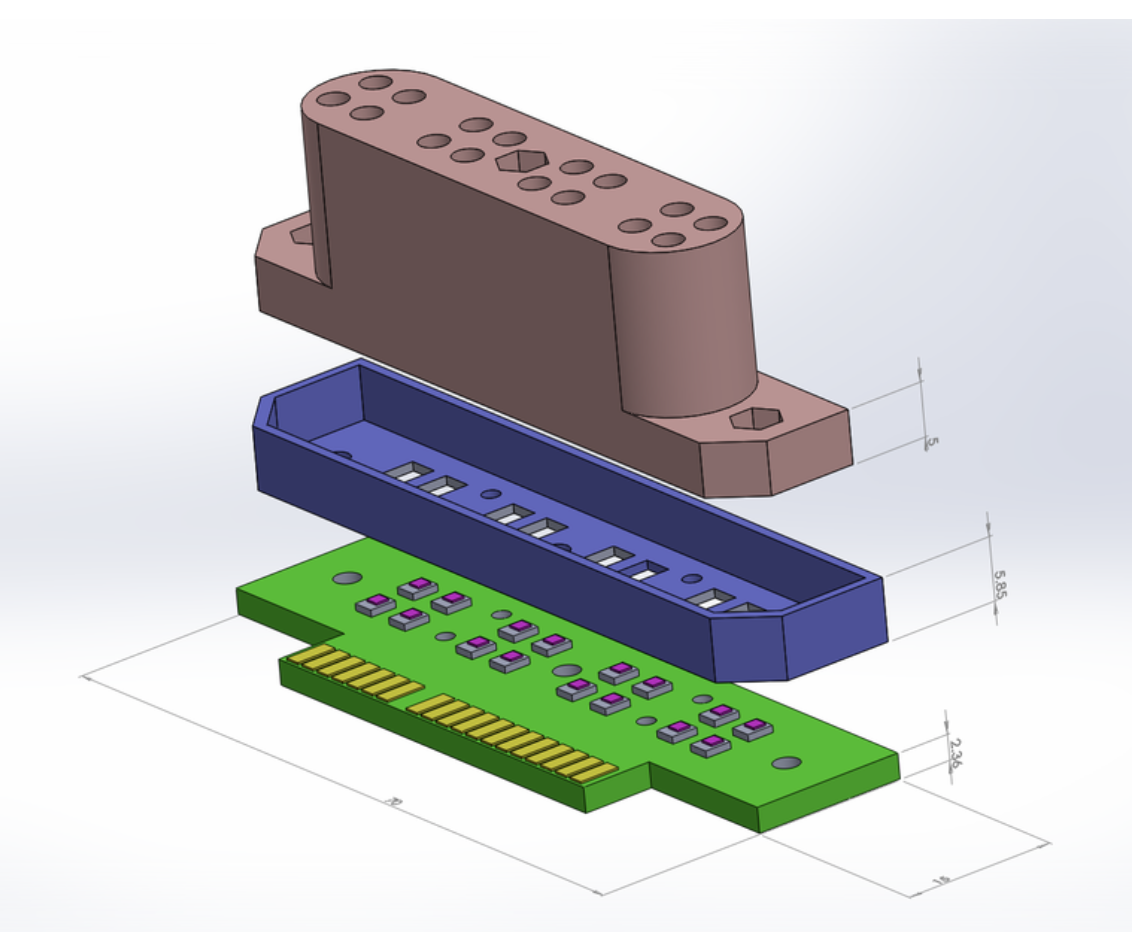}}
        \centering
        \ffigbox{\caption{Top: {\color{black}SiPM card with gold finger, to be inserted in the FEE board}, Bottom: SiPM spacer}\label{fig:SiPM Photo}}{
            \includegraphics[width=0.15\textwidth,angle=90]{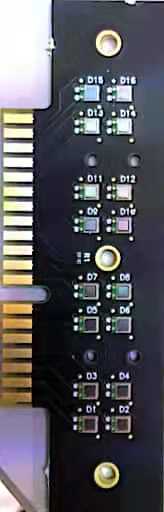}
            \includegraphics[width=0.1\textwidth,angle=90]{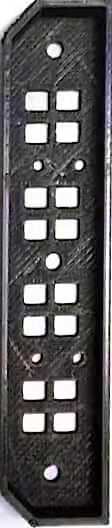}
            }
        \centering
    \end{floatrow}
\end{figure}

\begin{figure}[ht]
\centering
\mbox{\includegraphics[width=0.35\textwidth]{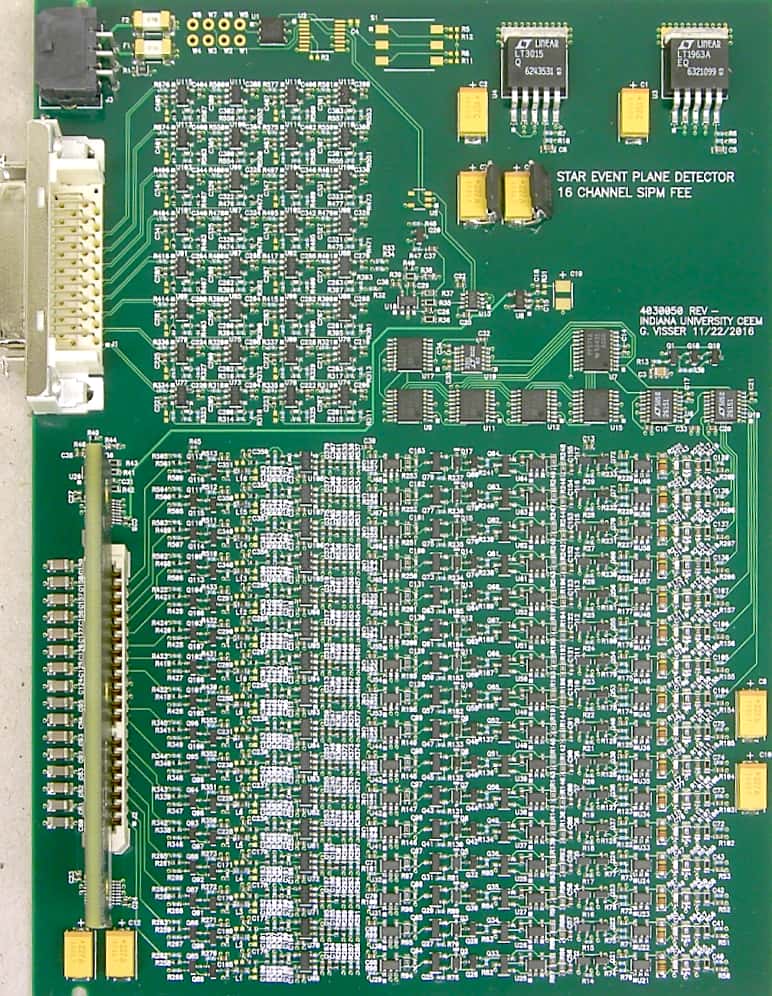}}
\mbox{\includegraphics[width=0.3\textwidth]{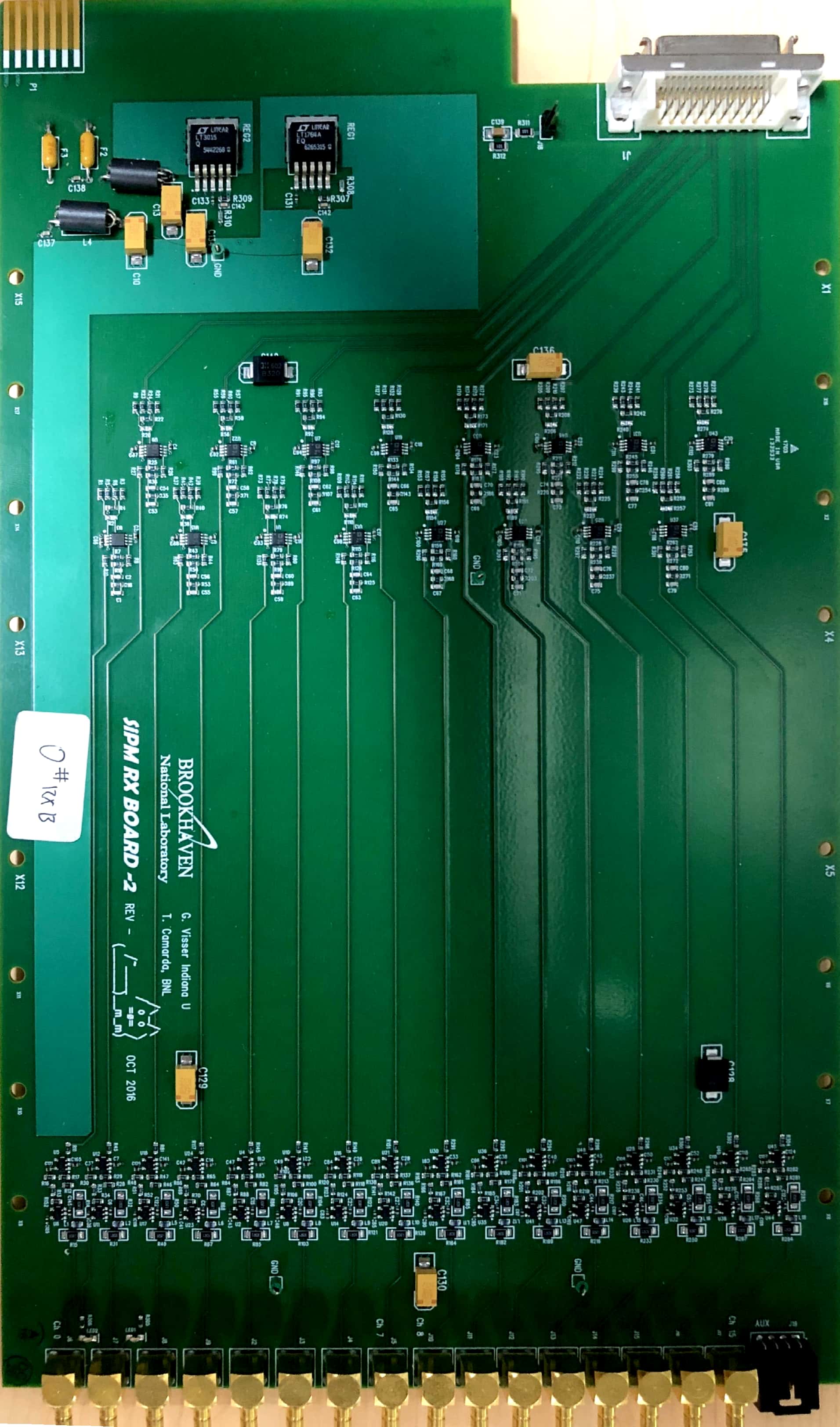}}
\caption{Left: FEE board; Right: RX board}
\label{fig:FeeAndRx}
\end{figure}

\section{Quality Assurance}
\label{sec:ElectronicsQA}

{\color{black} The SiPMs, FEEs and RX boards play a critical role in converting photons from the mounted SSs into clean electrical pulses for downstream digital conversion.}
During the production, all these items {\color{black} were} produced or manufactured by different companies: SiPMs by Hamamatsu photonics, FEE boards by E-TekNet Inc., RX and SiPM boards by Fastprint Circuit Tech. Co. Ltd, and SiPM assembly by Edadoc Co. Ltd.\footnote{Their websites are listed as below, Hamamatsu: https://www.hamamatsu.com/, E-TekNet: https://www.e-teckoffer.com, Fastprint: www.chinafastprint.com, Edadoc: http://www.edadoc.com/}.
Besides some basic quality checks were provided by the vendors, and additional checks {\color{black} were} required to ensure the quality of these products.

{\color{black} Given the channel count and EPD production schedule (accelerated to make full use of the RHIC beamtime),
approximately 1000 channels of the electronics chain needed to be produced and thoroughly tested in only four months.
Once the most important aspects of the QA were understood, we needed to perform the QA tests on complete SiPM-FEE-RX chains, which focused on the individual components only if a problem was found.}


First, we needed to ensure that SiPM could properly receive light from the clear fiber, which meant the air coupling between the fiber and SiPM needed to be correct.
This could be achieved by controlling the position of SiPM on the board and by controlling its height relative to the board surface.
To ensure uniform light coupling and to reduce the stress, the SiPMs were machine soldered onto the board with 0.05mm tolerance (horizontal and vertical).
Each SiPM board was individually examined under a microscope\footnote{\color{black} Optical Gaging Products (ogp) StarLite\textsuperscript{TM} 200.} to check if there were any flux residues or dust around the soldering point or if there were any visible scratches on the surface of SiPMs.
Most importantly, all the 16 SiPMs needed to be soldered firmly and at the right positions; the height and parallelism of each SiPM was measured and controlled, as shown in Figure~\ref{fig:SiPM Visual Check}.
{\color{black}
The microscope is able to measure the distance between different points. Practically we chose 3 couples of points along the edge of SiPM and the PCB surface respectively, and the microscope gave the distances between each couple of points. The average of the measurements is taken as the height of SiPM, the resolution of which is about 10$\mu$m and is enough to guarantee the light tightness of the fiber-SiPM coupling.
}

\begin{figure}[ht]
  \centering
  \mbox{\includegraphics[width=0.35\textwidth]{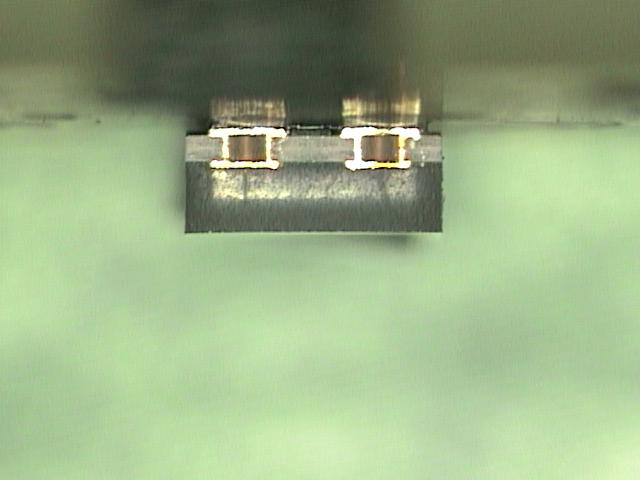}}
  \mbox{\includegraphics[width=0.35\textwidth]{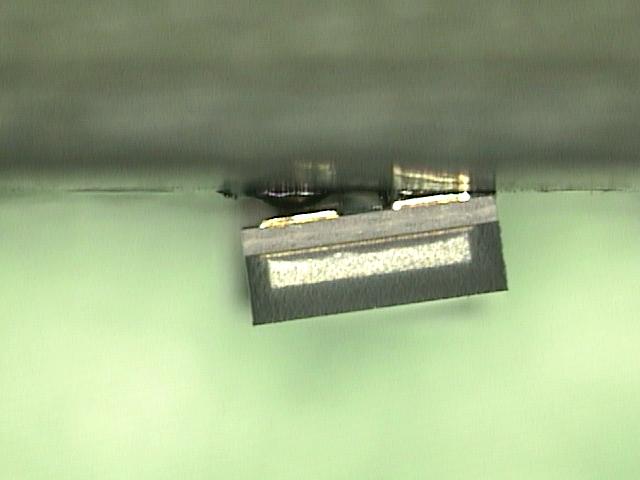}}
  \caption{Left: SiPM assembled properly. Right: SiPM assembled badly. We choose three points on the SiPM surface and the average distances to the surface of PCB board as the measured height.}
  \label{fig:SiPM Visual Check}
\end{figure}

Second, each SiPM needs to function properly with the expected characteristics, such as gain, noise, etc.

This {\color{black}preliminarily} requires the FEE board to correctly communicate with the data acquisition (DAQ) devices, provide HV for SiPM with controllable value, and transmit signals from SiPM and process it as desired.
{\color{black}After inserting the SiPM board to the FEE board, we set different bias to measure dark current through current monitor on FEE board and acquire noise spectrum of SiPM from oscilloscope, in order to check the basic connection and functions of the SiPM.}


Gain is one of the most important parameters of SiPM and is influenced by the bias HV and temperature.
It is primarily determined by the overvoltage, which is defined as the quantity of supplied bias voltage beyond the breakdown voltage, $V_{br}$.
The breakdown voltage is the bias voltage at which the electric field strength in the depletion region of the SiPM is sufficient to create a Geiger discharge.
Also, $V_{br}$ should also be the bias voltage at which the electrons in the SiPM stop multiplication during the avalanche procedure.
Thus, the overvoltage determines how much charge is produced during an avalanche.
{\color{black} The signal charge is calculated by the simple formula $Q = C \cdot \Delta V$, where $C$ is the capacitance of SiPM and $\Delta V$ is the overvoltage.}
The gain of SiPM, which is defined as the output charge of a single photoelectron (SPE), increases linearly with the bias voltage under a fixed temperature ~\cite{2002atpp.conf..717B}.

We have tried two methods to determine $V_{br}$.
One was to measure the voltage-current (U-I) curve; another was to directly acquire the Gain-$\Delta V$ relationship.
In the UI curve, where U is the supplied voltage, and I is the dark current of SiPM, the current rises rapidly at $V_{br}$.
Thus, the UI curve, as shown in Figure~\ref{fig:UI_Curve}, can be used to measure $V_{br}$.
However, different theories will give different values for $V_{br}$ ~\cite{NAGY201755}.
Because of the relatively low precision (0.01 $\mu$A) of the current monitor on the FEE board, 
{\color{black}
we choose the inflection point of UI curve as breakdown threshold, the corresponding abscissa of the intersection point is determined as breakdown voltage, $V_{br}$, and calculate $V_{op} = V_{br} + 5$ V.
}

For a Gain-$\Delta V$ line, the $V_{br}$ can be measured by finding the intersection point with $Gain=0$ line.
Figure~\ref{fig:Gain_Fit} shows a charge spectrum of SiPM at 38$^\circ C$, measured by a charge-to-digital converter (QDC) (CAEN V965A, 25fC/ch).
A multi-Gaussian fit was applied to the data, wherein the first peak was the pedestal, and the other peak represented single and multiple photoelectrons.
The distance between the adjacent two peaks is the gain of one photoelectron. 
Figure~\ref{fig:BD_Fit} depicts how the extracted gain of SiPM increases linearly with the bias voltage, as expected.
The bias voltage at which the fit line crosses the X-axis is $V_{br}$. 

Results from both the methods provide similar $V_{br}$ values.
Although the value from the U-I curve has a lower resolution than that from the Gain-$\Delta V$ line, the U-I curve is much easier to measure, and the resolution is sufficient for our QA.
Thus, we finally choose to measure the UI curve to get $V_{br}$ for each SiPM.
\begin{figure}[ht]
  \centering
  \subfloat[U-I curve in the Cartesian coordinate system.]{\includegraphics[width=0.45\textwidth]{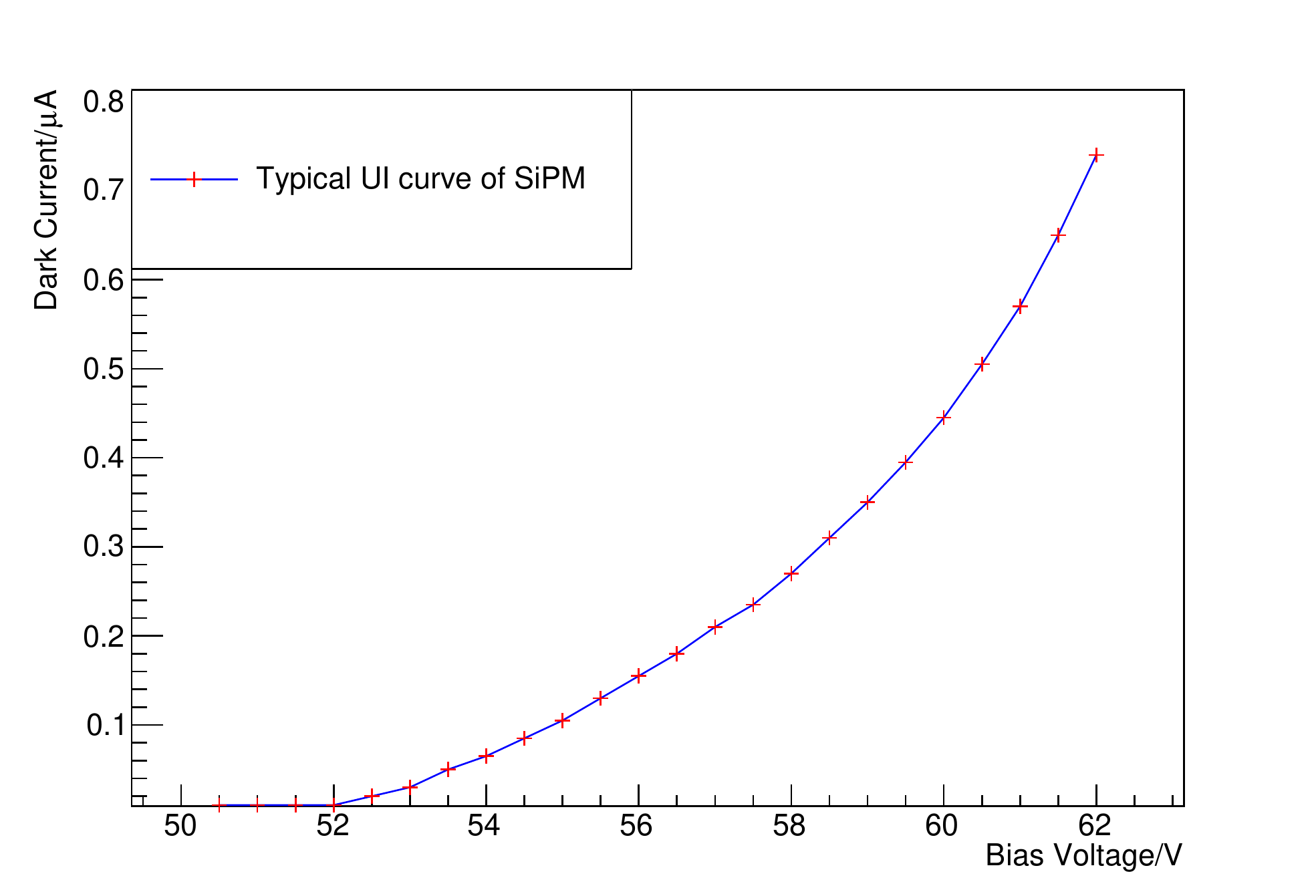}}\hspace{20pt}
  \subfloat[U-I curve in the Logarithm coordinate system.]{\includegraphics[width=0.45\textwidth]{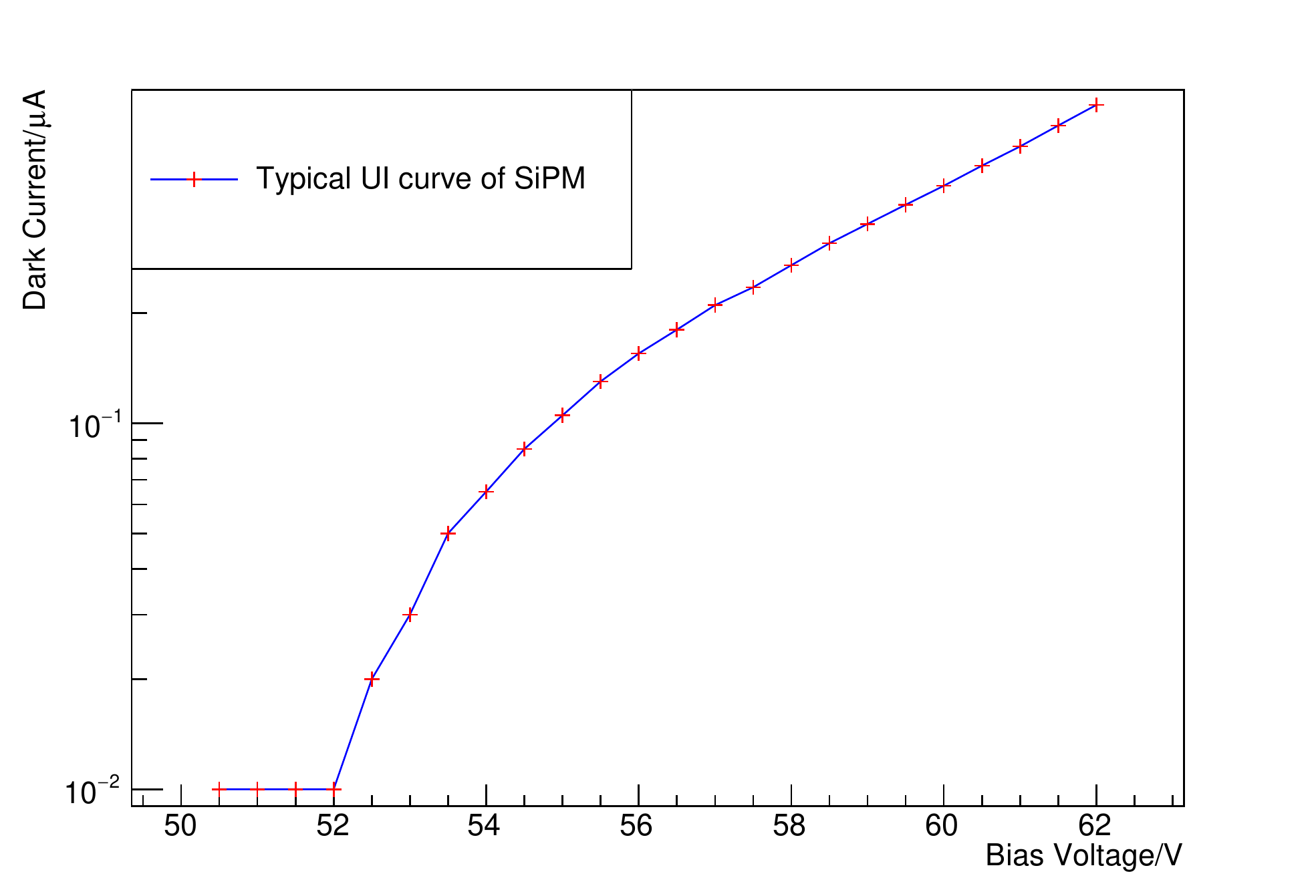}}\
  \caption{U-I Curve}\label{fig:UI_Curve}
\end{figure}

\begin{figure}[ht]
  \centering
  \subfloat[Charge spectrum of SiPM]{\includegraphics[width=0.45\textwidth]{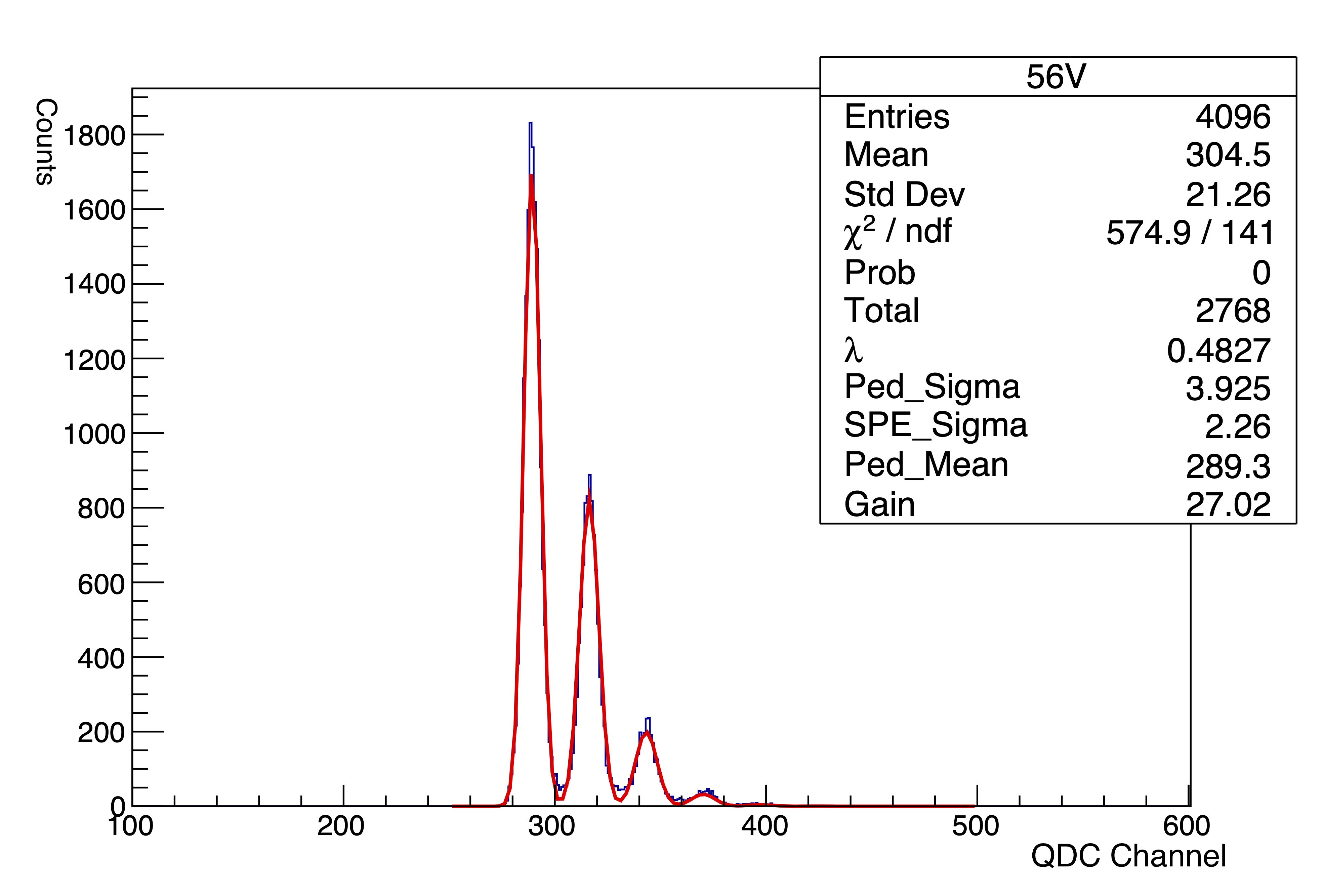}\hspace{20pt}\label{fig:Gain_Fit}}
  \subfloat[Gain versus bias voltage fit with a linear function]{\includegraphics[width=0.45\textwidth]{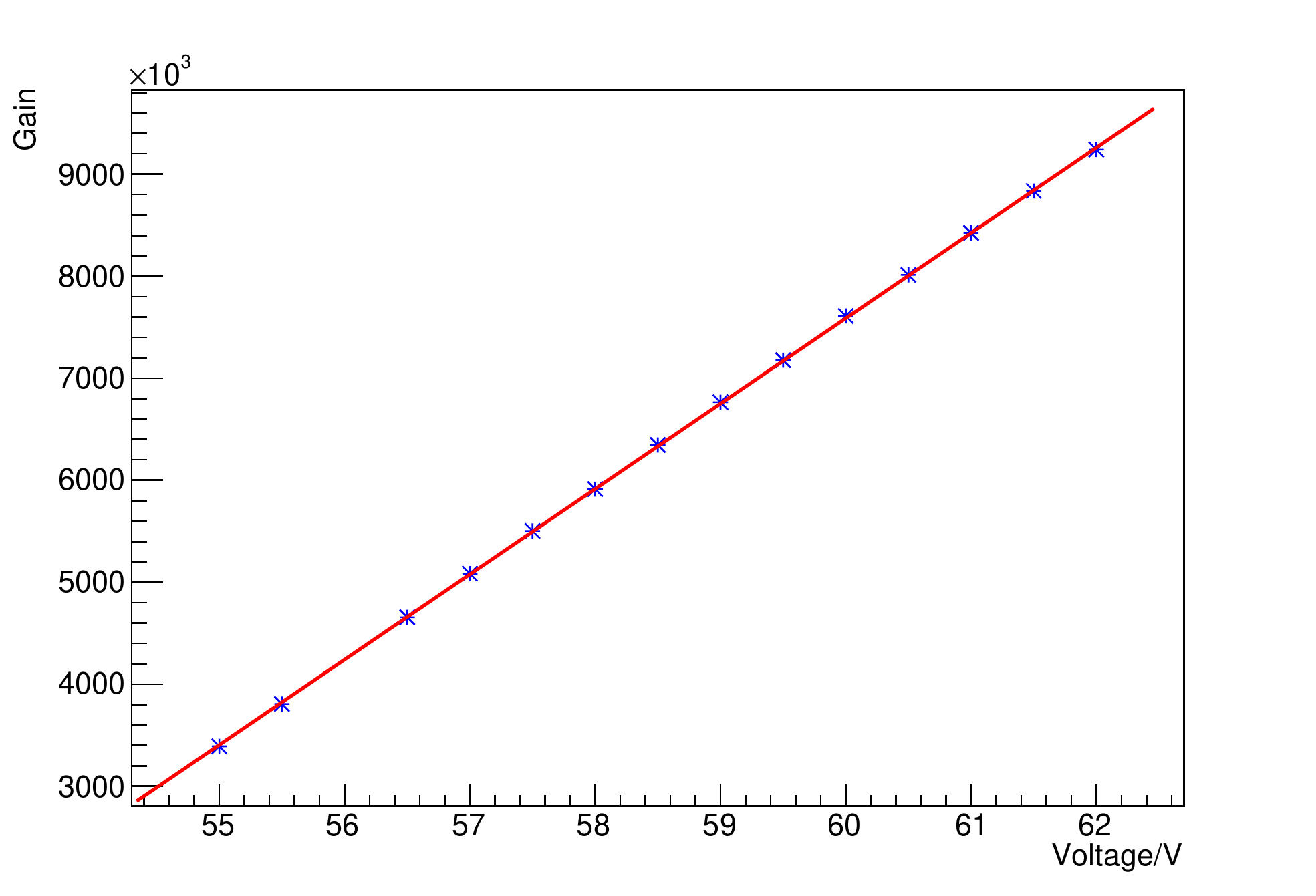}\label{fig:BD_Fit}}
  \caption{Breakdown voltage measured via gain variations. See text for details.}
  \label{fig:Gain_Test}
\end{figure}

$V_{br}$ of SiPM usually increases for higher temperature~\cite{4774854}.
Therefore the environment temperature will also influence the gain of SiPM. The {\color{black} thermistor} on the SiPM board can sense the local temperature change, which is used (by the FEE) to compensate for the gain loss accordingly by adjusting the bias voltage.
To keep the SiPM operation stable at different temperature conditions, the effect of gain compensation needs to be tested.
The expected compensation slope provided by the vendor is approximately 54 $\rm mV/^\circ C$, as shown in Figure~\ref{fig:Incomplete_Compensation}.
Compared with the measurement without temperature compensation, the dependence of the gain on the temperature was greatly suppressed, which also demonstrates the need for such corrections.
It was also found that the gain variation with temperature in Figure~\ref{fig:Incomplete_Compensation} had not been fully corrected; {\color{black} this was because of the temperature dependence of the input voltage of the common-base preamplifier.}
In our test, we calculated the correct compensation slope parameter as 62.5 $\rm mV/^\circ C$, the compensation effect under which is shown in Figure~\ref{fig:Complete_Compensation}. In our QA procedure the temperature compensation parameter 62.5 $\rm mV/^\circ C$ was used and under this condition, the operation voltage of each SiPM needed to fall into the 50 V to 60 V interval.

\begin{figure}[ht]
  \subfloat[No compensation versus incomplete compensation]{\includegraphics[width=0.45\textwidth]{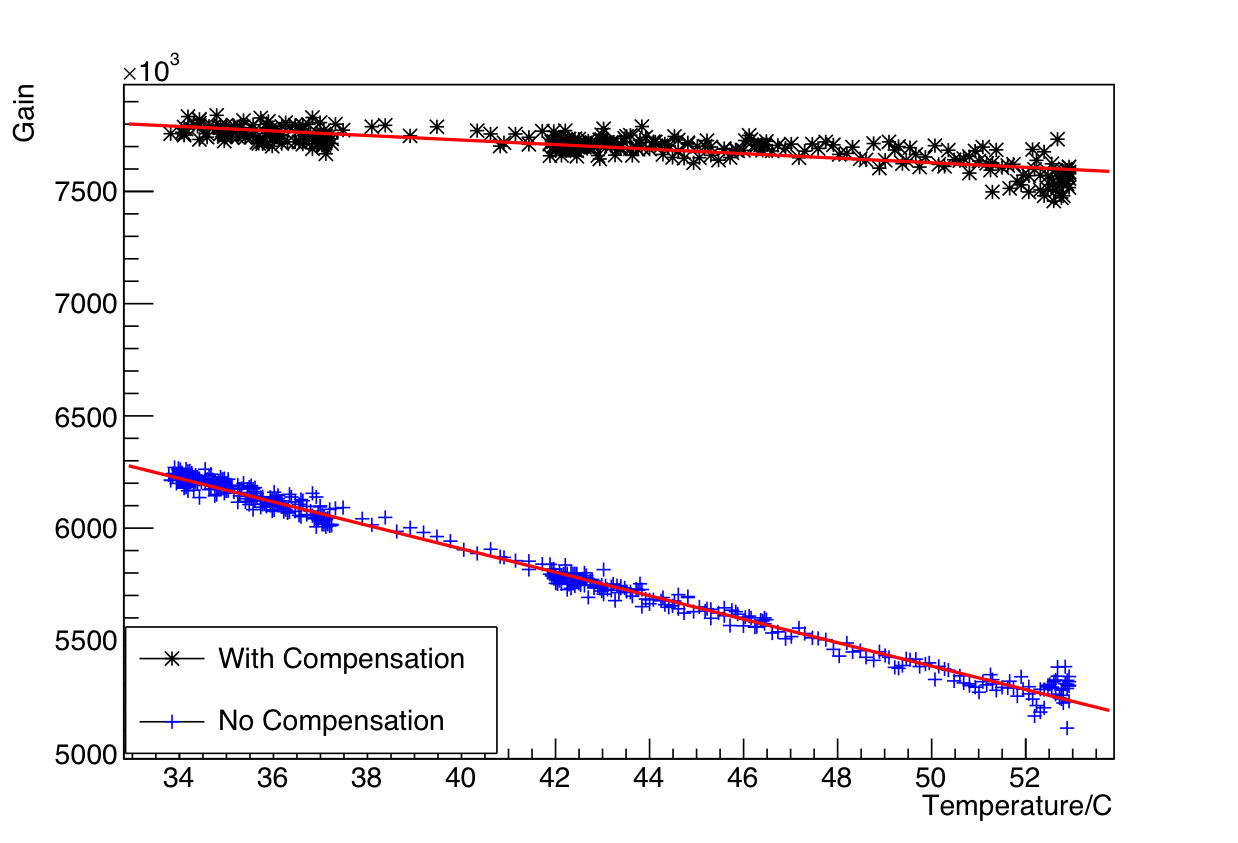}\hspace{20pt}
  \label{fig:Incomplete_Compensation}}
  \subfloat[Incomplete compensation versus complete compensation]{\includegraphics[width=0.45\textwidth]{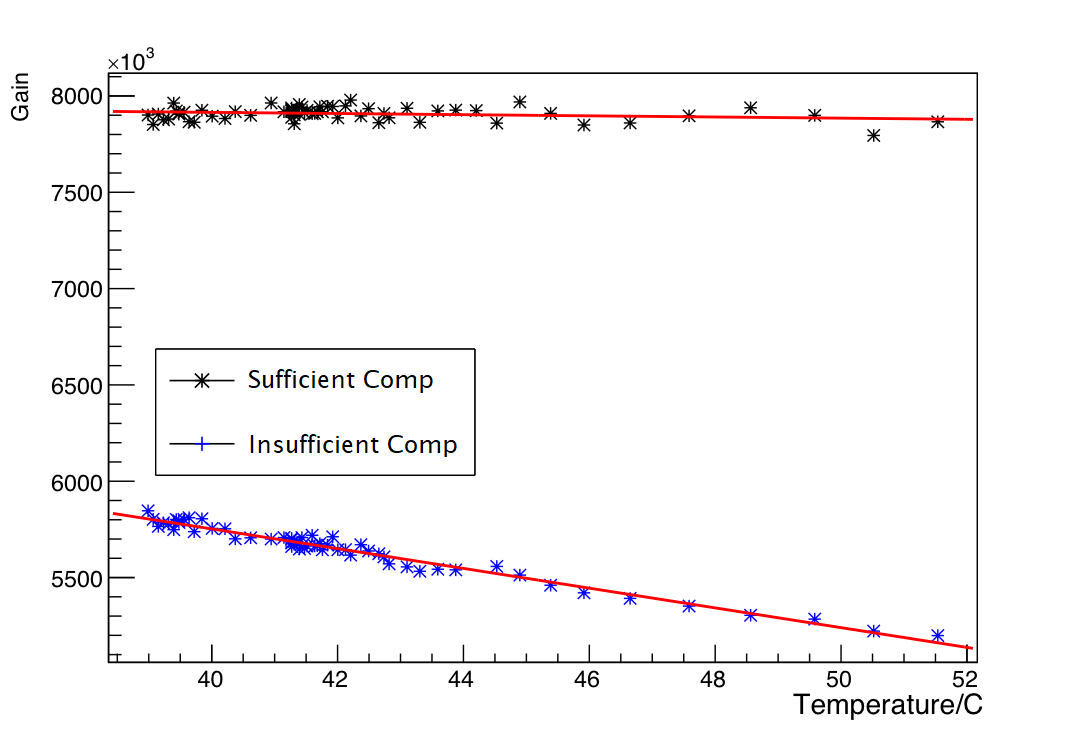}
  \label{fig:Complete_Compensation}}
  \caption{Temperature compensation}
  \label{fig:Compensation}
\end{figure}

Energy resolution is another important characteristic of SiPM, especially for the counting of photons.
However the light yield is expected to be approximately 100 photoelectrons for a MIP hitting the EPD, and the fluctuation is huge (Landau fluctuation).
Therefore, energy resolution is not a crucial parameter in our research.
The typical oscilloscope signal (in the trace mode) for a single photoelectron (SPE) is shown in Figure~\ref{fig:SPE_OSC}.
Signals for multiple photons are also shown, and the splitting of waveforms is evident.
By fitting the charge spectrum measured by the QDC, as in Figure~\ref{fig:Gain_Fit}, the energy resolution (or the signal-to-noise ratio) can be quantitatively determined.
The charge spectrum can also be obtained by directly analyzing the recorded waveforms.
To save on measurement time, we use a 16-channel digitizer, instead of the oscilloscope, to record the signal waveform. Results from both the oscilloscope and the digitizer are shown in Figure~\ref{fig:SPE}.
Although the resolution of SPE is a little poorer with the digitizer than the oscilloscope, it is sufficient for our QA.
We define the SPE energy resolution as $R_{\rm{SPE}}=\sigma_{\rm{SPE}}/(\mu_{\rm{SPE}}-\mu_{\rm{PED}})$, where $\sigma_{\rm{SPE}}$ is the Gaussian fit $\sigma$ of SPE; $\mu_{SPE}$ and ${\mu_{PED}}$ are the mean values of the SPE and pedestal peaks respectively.
Note that the resolutions of SPE and pedestal are similar, which shows an unstable baseline as the main source of fluctuation.
{\color{black}
For our purpose $\sim5\%$ contamination of the SPE peak from pedestal is a satisfactory criteria which conservatively imply $R\lesssim0.4$($\sim2.5 \sigma$ separation between the two Gaussian peaks).
}

\begin{figure}[ht]
  \centering
  \subfloat[Single photon signal from the oscilloscope]{\includegraphics[width=0.45\textwidth]{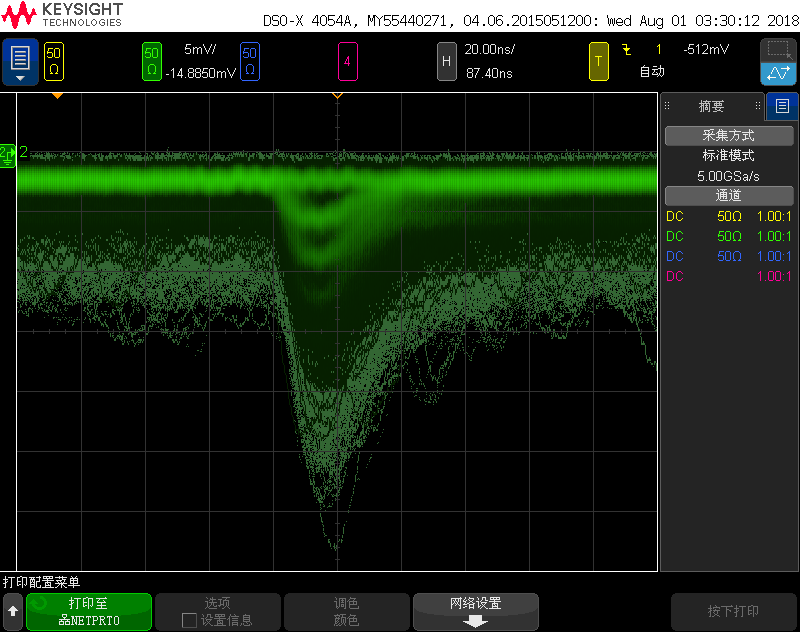}\label{fig:SPE_OSC}}
  \subfloat[Single photon signal from the digitizer; a small slope of the baseline is visible in the waveform]{\includegraphics[width=0.45\textwidth]{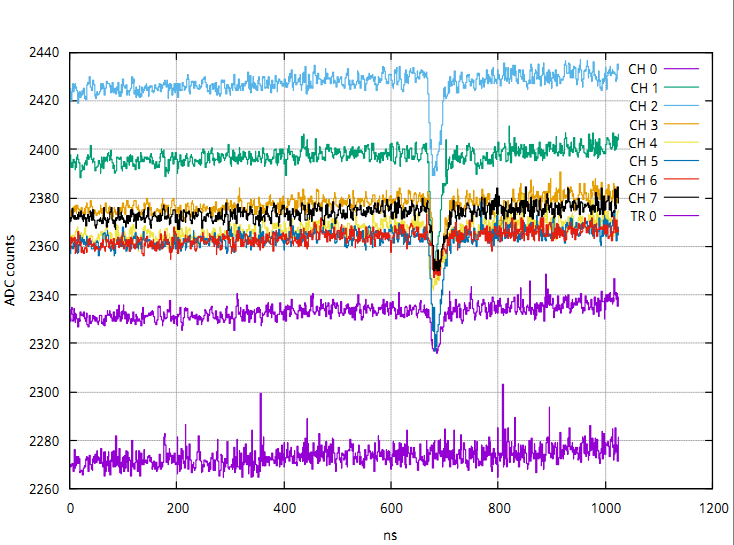}\label{fig:SPE_DIG}}\
  \subfloat[Spectrum from the oscilloscope]{\includegraphics[width=0.45\textwidth]{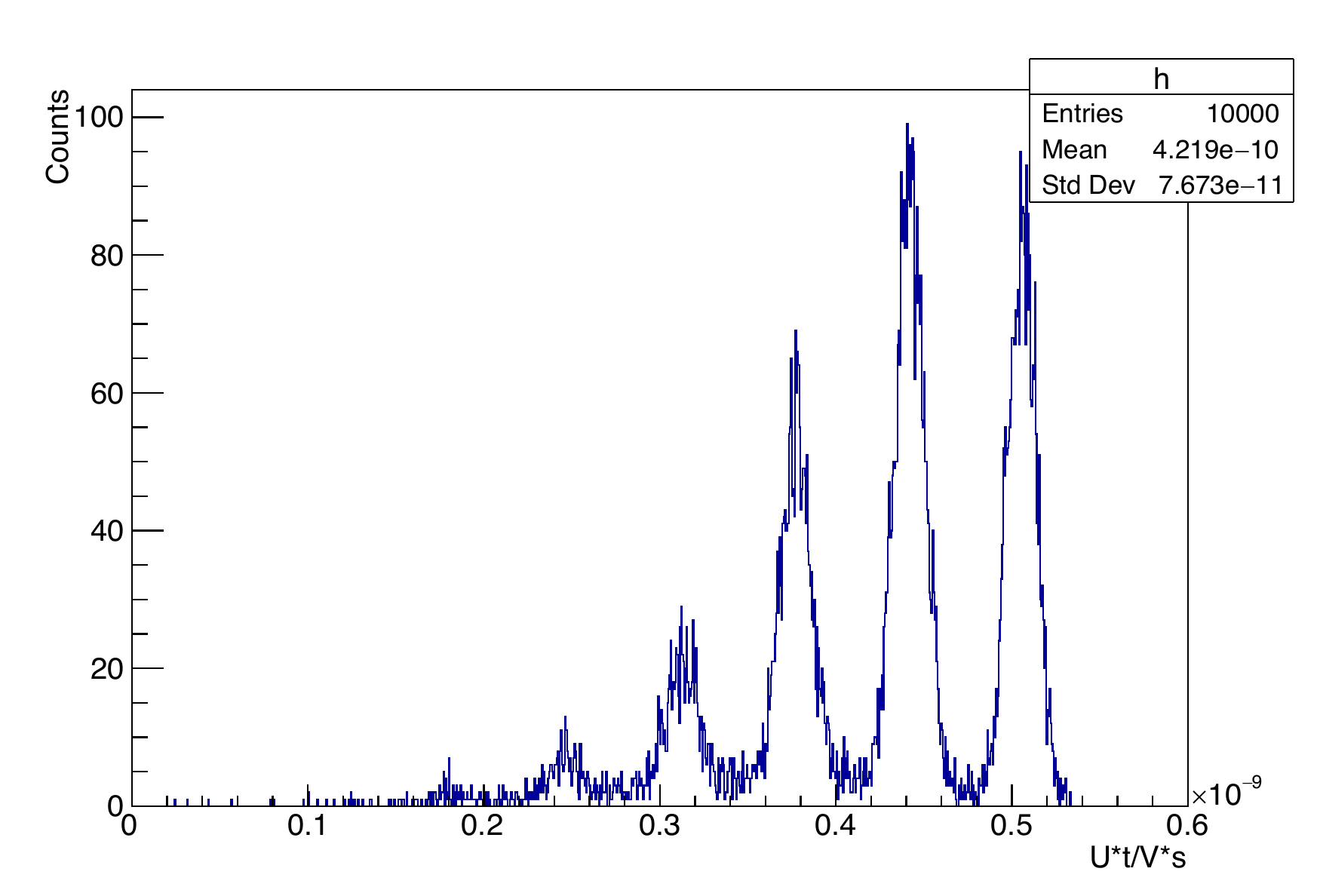}\label{fig:SPE_OSC_Spectrum}}
  \subfloat[Spectrum from the digitizer, with the background subtracted]{\includegraphics[width=0.45\textwidth]{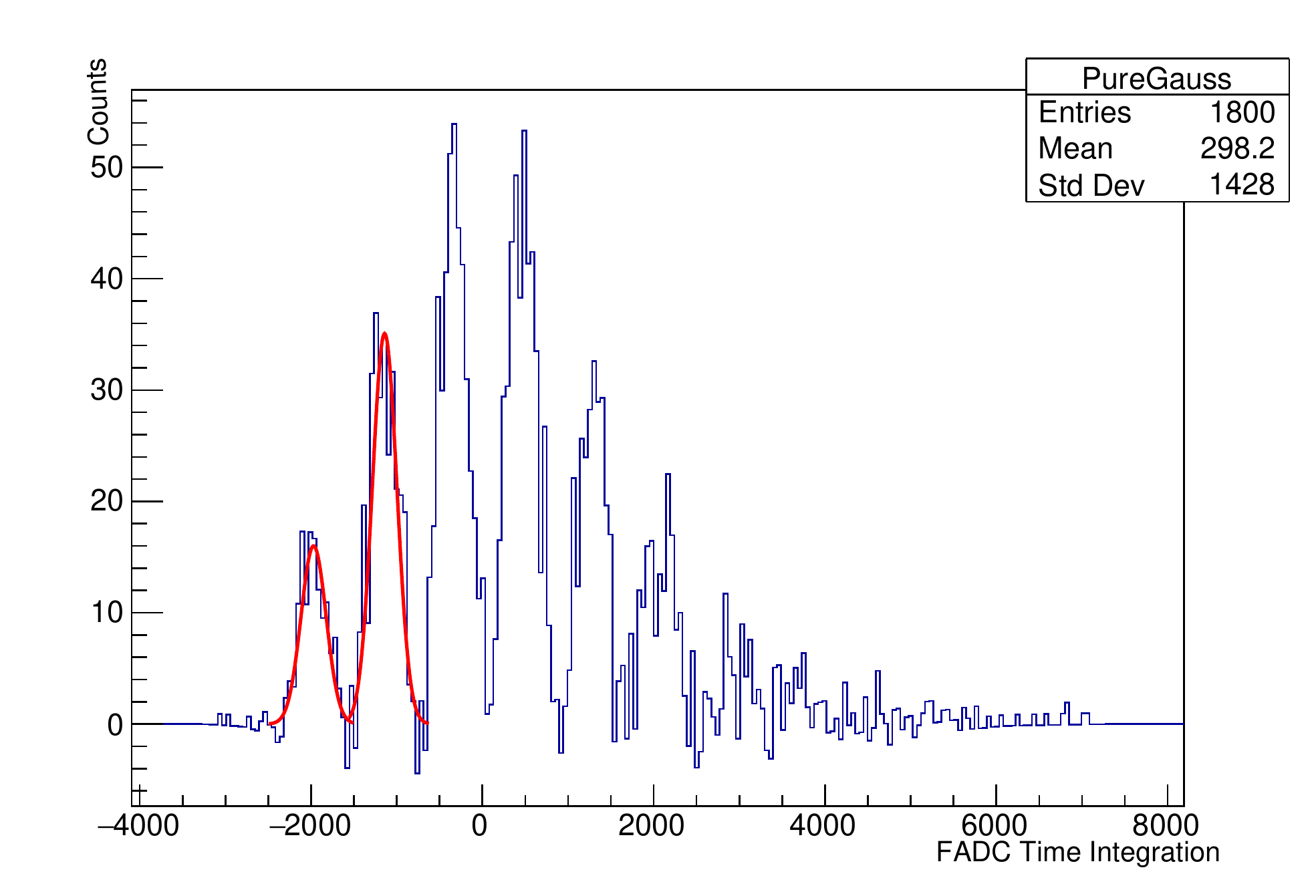}\label{fig:signal-dgtz}}
  \caption{Single photon signal}
  \label{fig:SPE}
\end{figure}

Besides maintaining the proper operation of SiPM, the FEE board can also provide {\color{black} adjustable pedestals for each channel.}
In our QA, the range of the pedestal needs to contain a zero value.


The last QA item is to check if unwanted noise was introduced into the signal after being shaped and filtered by the electronic boards at the front end.
By analyzing the frequency spectrum of the signals and noises, we can check whether any particular noise patterns appear.
When measuring the noise spectrum of a set of validated reference boards (SiPM board, FEE board and RX board), the typical frequency spectrum is smoothly continuous (see Figure~\ref{fig:Digitizer_Noise}), without any notable sharp peaks or spikes.
Both the oscilloscope and digitizer were used to acquire the noise spectrum.
From Figure~\ref{fig:Oscilloscope_Noise}, we can see that the frequency spectrum measured by the oscilloscope shows clear spikes at certain frequencies.
However, these spikes can be seen as features of our oscilloscope, because they exist even without any inputs, as in Figure~\ref{fig:Oscilloscope_Bottom}.
The measurement done by using the CAEN N6742 digitizer is shown in Figure~\ref{fig:Digitizer_Noise}.
Although the result obtained from the digitizer is less accurate than that obtained from the oscilloscope, it does not show any specific noise patterns.
We choose the digitizer in our QA to check the noise frequency spectrum, in which no obvious peaks or differences from the reference spectrum are allowed.
\begin{figure}[ht]
  \centering
  \subfloat[Electronic noise in oscilloscope]{\includegraphics[width=0.45\textwidth]{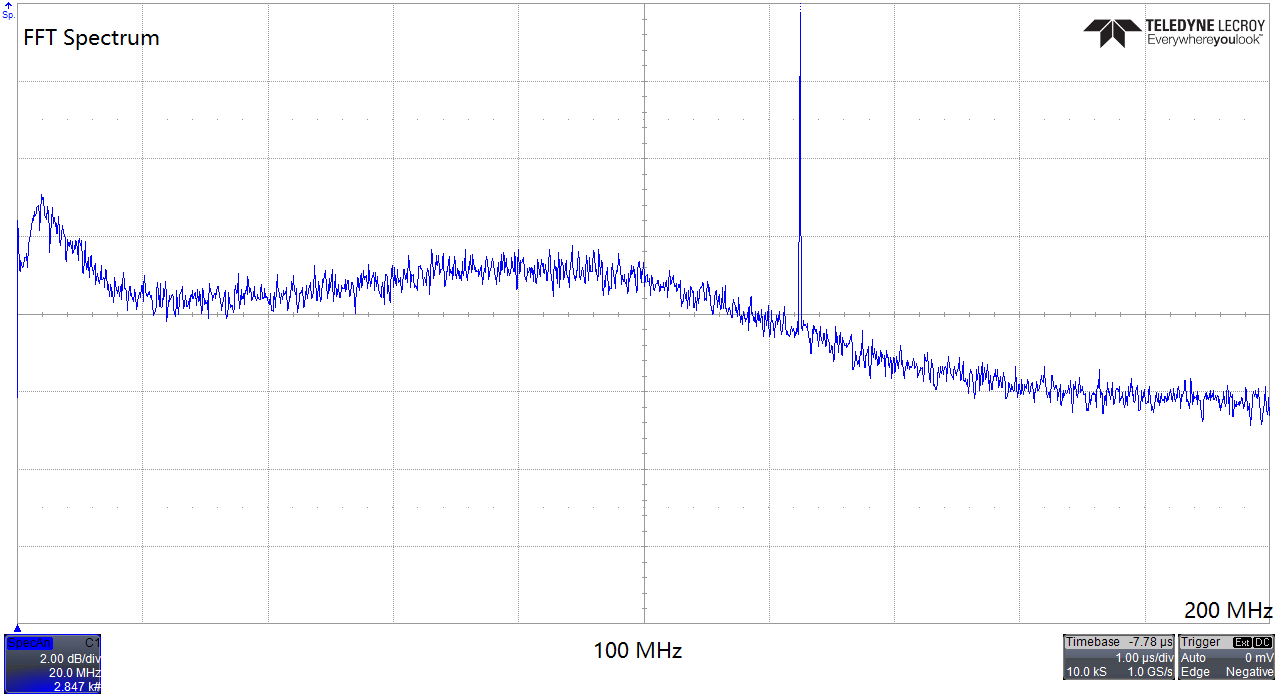}\label{fig:Oscilloscope_Noise}}\hspace{20pt}
  \subfloat[Oscilloscope noise with no input signal]{\includegraphics[width=0.45\textwidth]{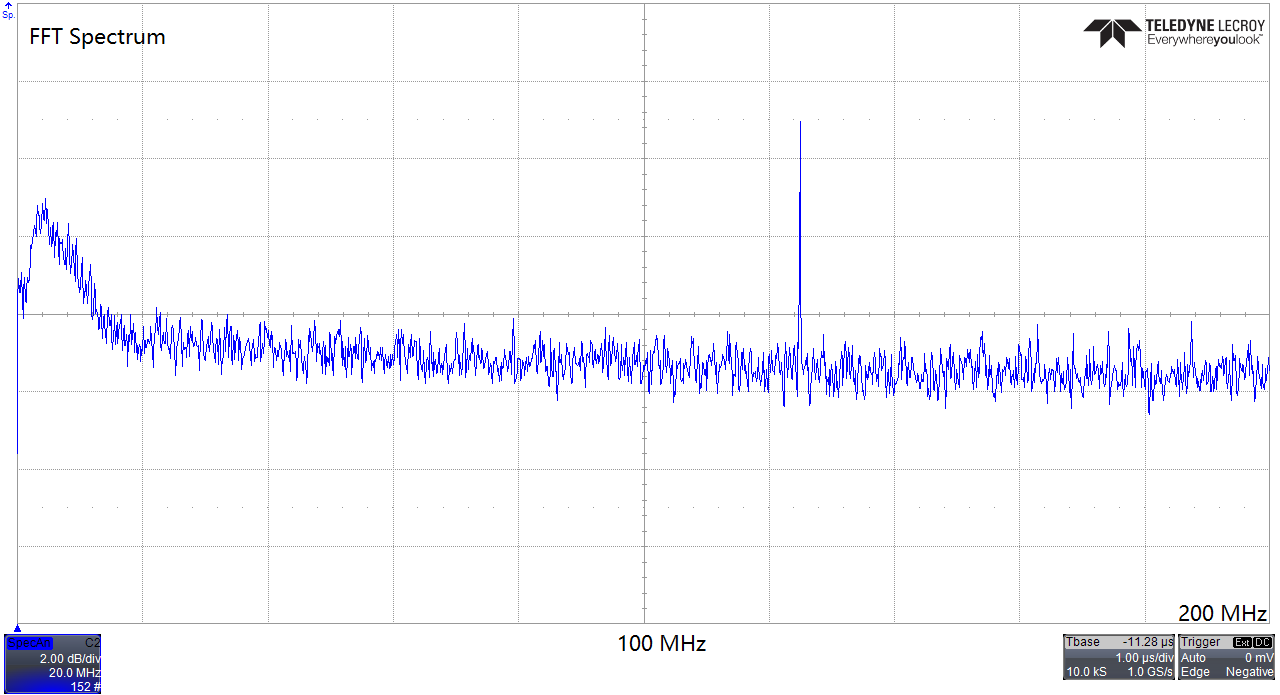}\label{fig:Oscilloscope_Bottom}}\
  \subfloat[Electronic noise in digitizer]{\includegraphics[width=0.5\textwidth]{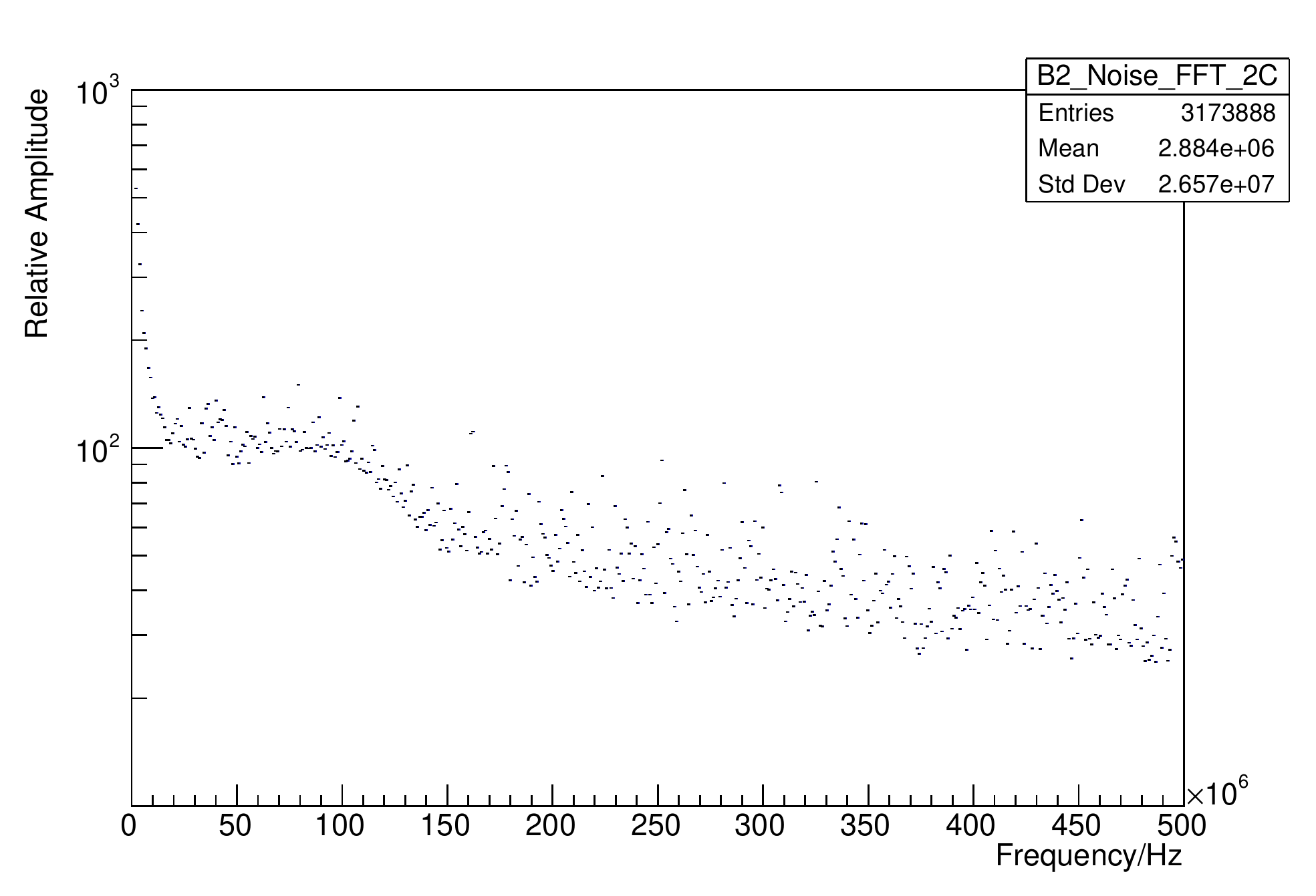}\label{fig:Digitizer_Noise}}
  \caption{Comparison of electronic noise between digitizer \& oscilloscope}\label{fig:Noise_Comparison}
\end{figure}

Note that the timing performance is not included in the QA procedure. 
{
\color{black}
The production QA only detects production problems; therefore, most kinds of defects (e.g., bad solder joint, wrong component value, etc.) would already have been flagged by the tests conducted.
Meanwhile, the timing performance of the EPD is limited by the scintillator and light collection than by the electronics.
}

\begin{figure}[ht]
  \centering
  \begin{floatrow}
  \centering
  \ffigbox{\caption{System Flow Chart}}{\includegraphics[width=0.48\textwidth]{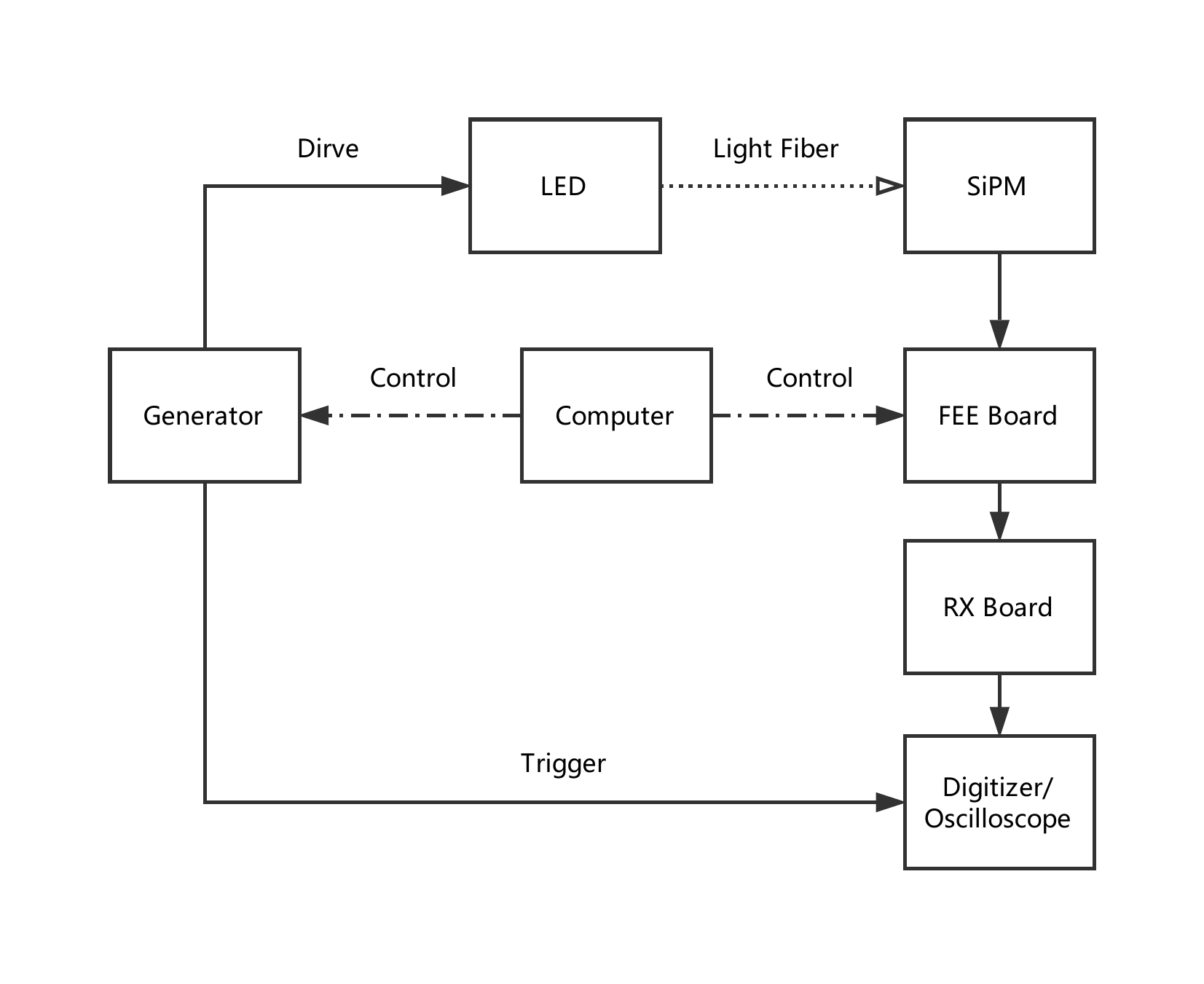}\label{fig:System_Chart}}
  \centering
  \ffigbox{\caption{Light source system}\label{fig:Light_Source}}{
    \includegraphics[width=0.35\textwidth]{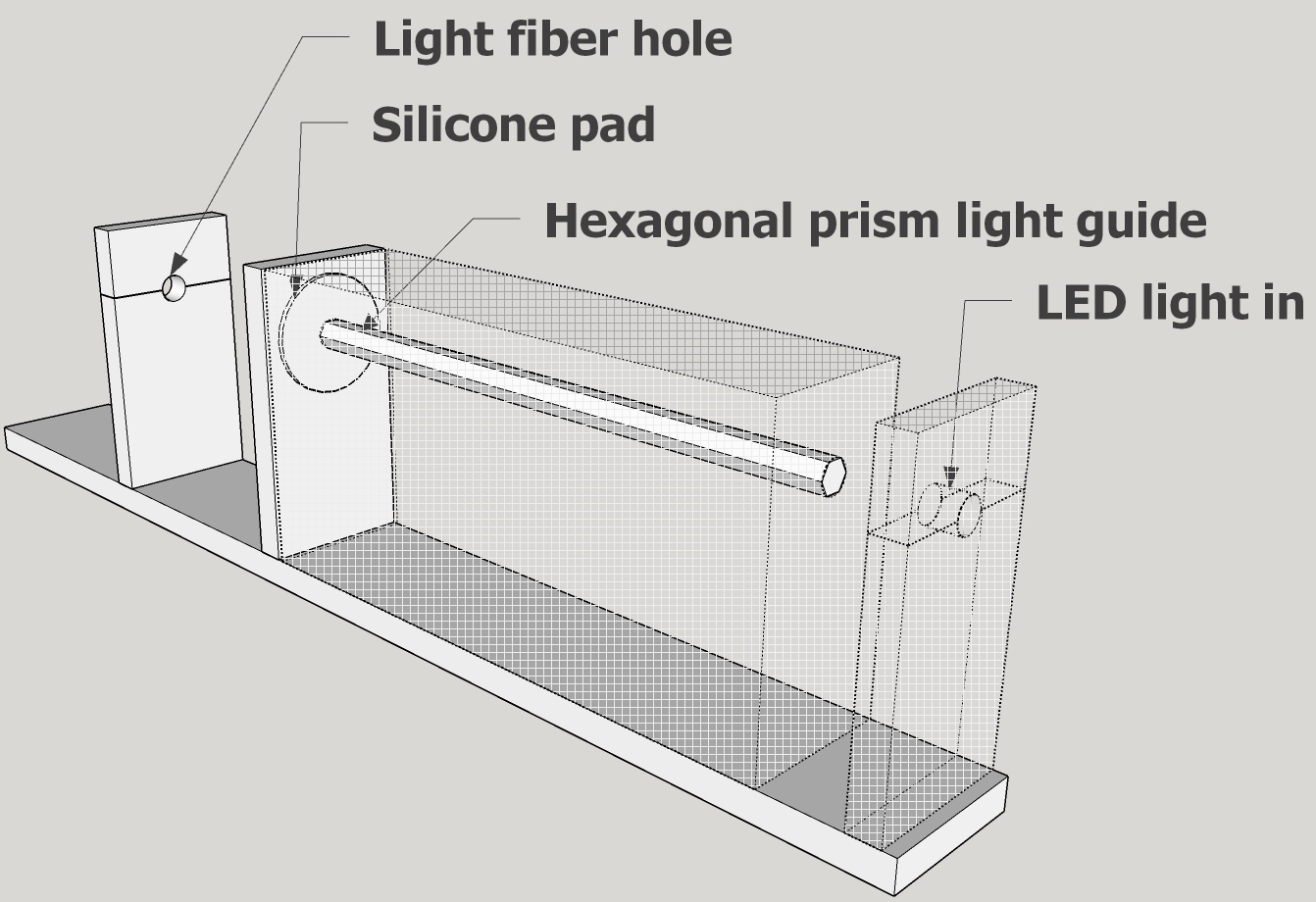}\\
    \includegraphics[width=0.35\textwidth]{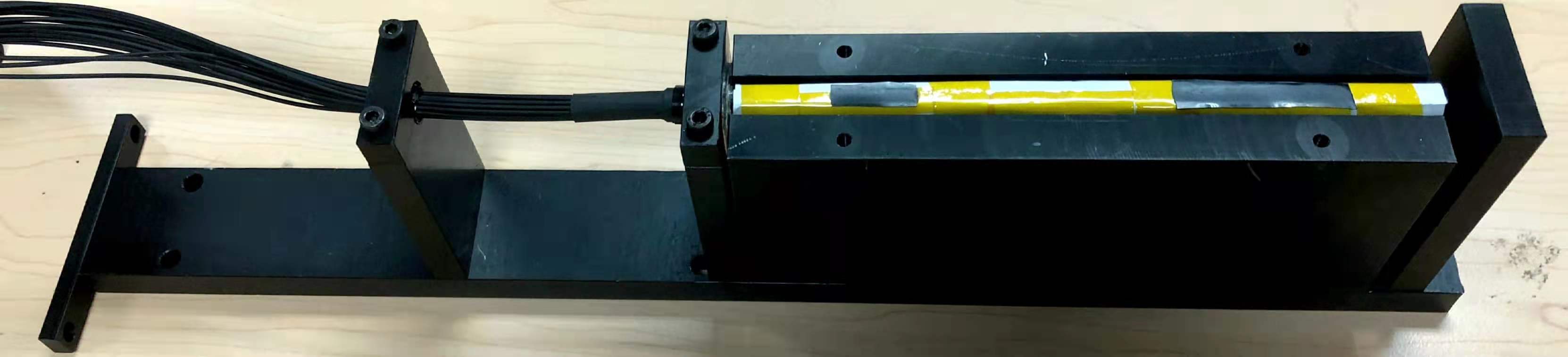}
    }
  \end{floatrow}
  \end{figure}


  \begin{figure}[ht]
    \begin{floatrow}
    \centering
    \subfloat[Waveforms obtainded while measuring the pedestal. When the bias was set to 46V, DAC at 127, 0 and -127, the waveforms showed the pedestal adjustable range.]{\includegraphics[width=0.44\textwidth]{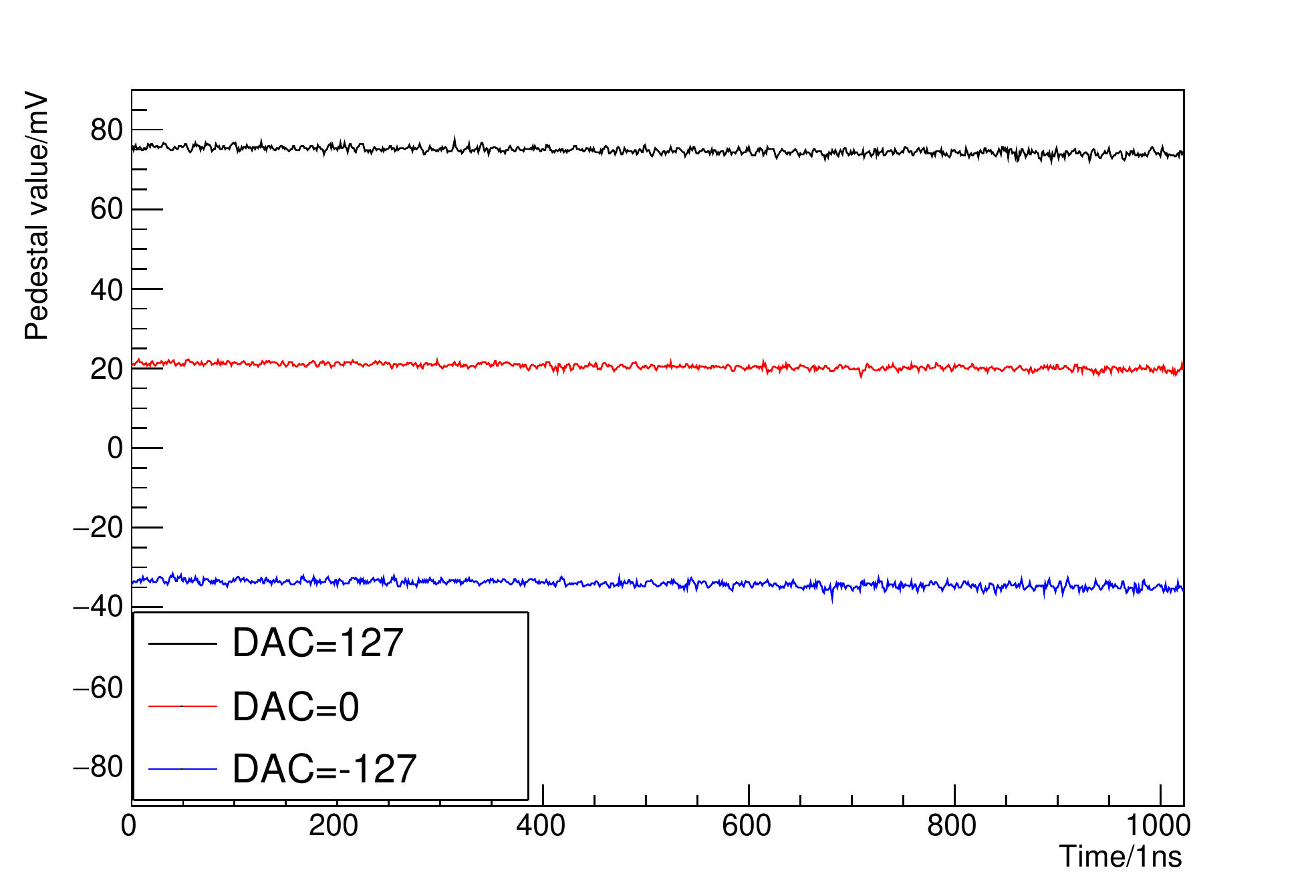}\label{fig:waveform_DAC}}
    \subfloat[Pedestal vs. DAC setting. A similar tendency and almost the same slope can be seen from the Figure.]{\includegraphics[width=0.44\textwidth]{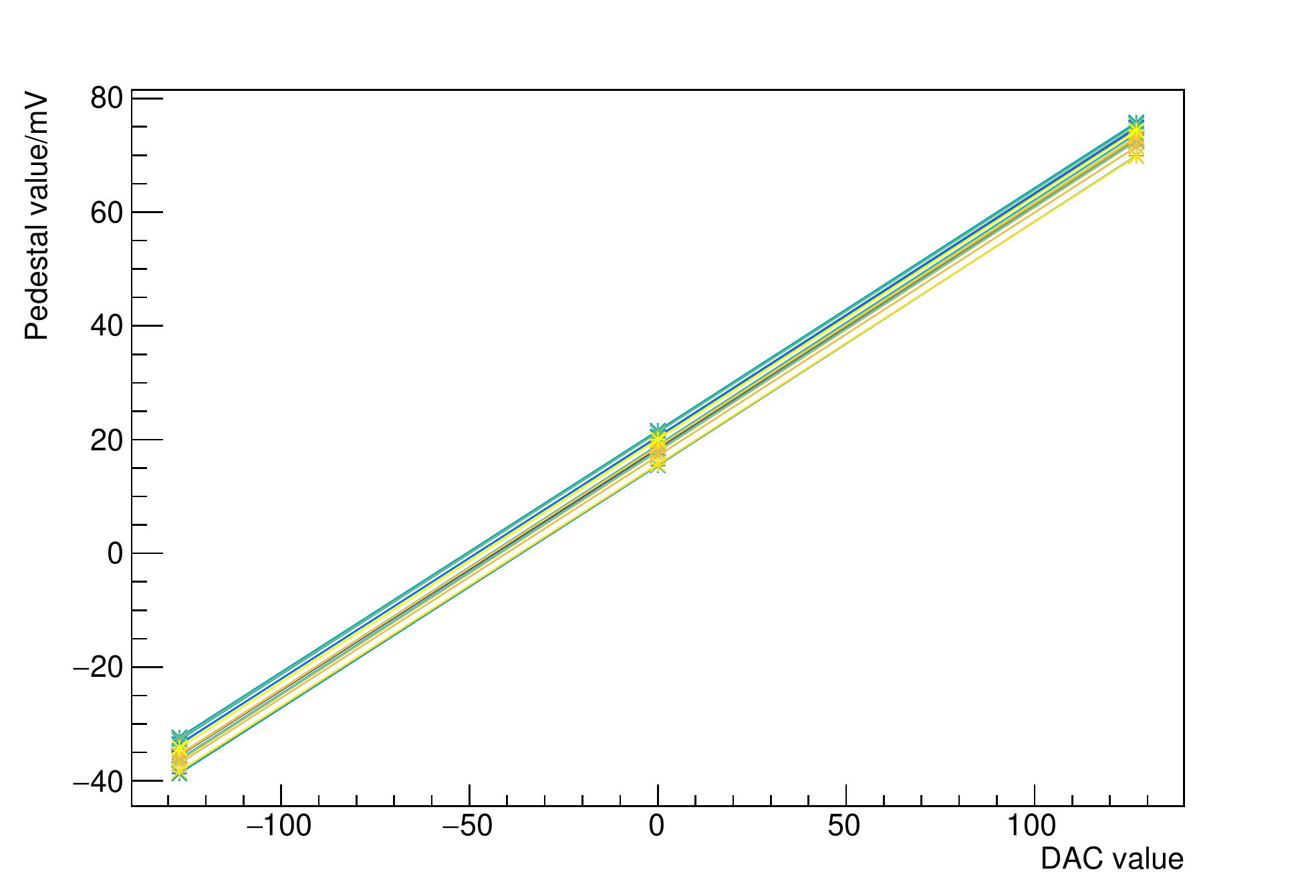}\label{fig:ped_DAC}}
    \caption{DAC measurement.}\label{fig:DAC}
    \end{floatrow}
  \end{figure}

  \begin{figure}
    \centering
    \includegraphics[width=0.45\textwidth]{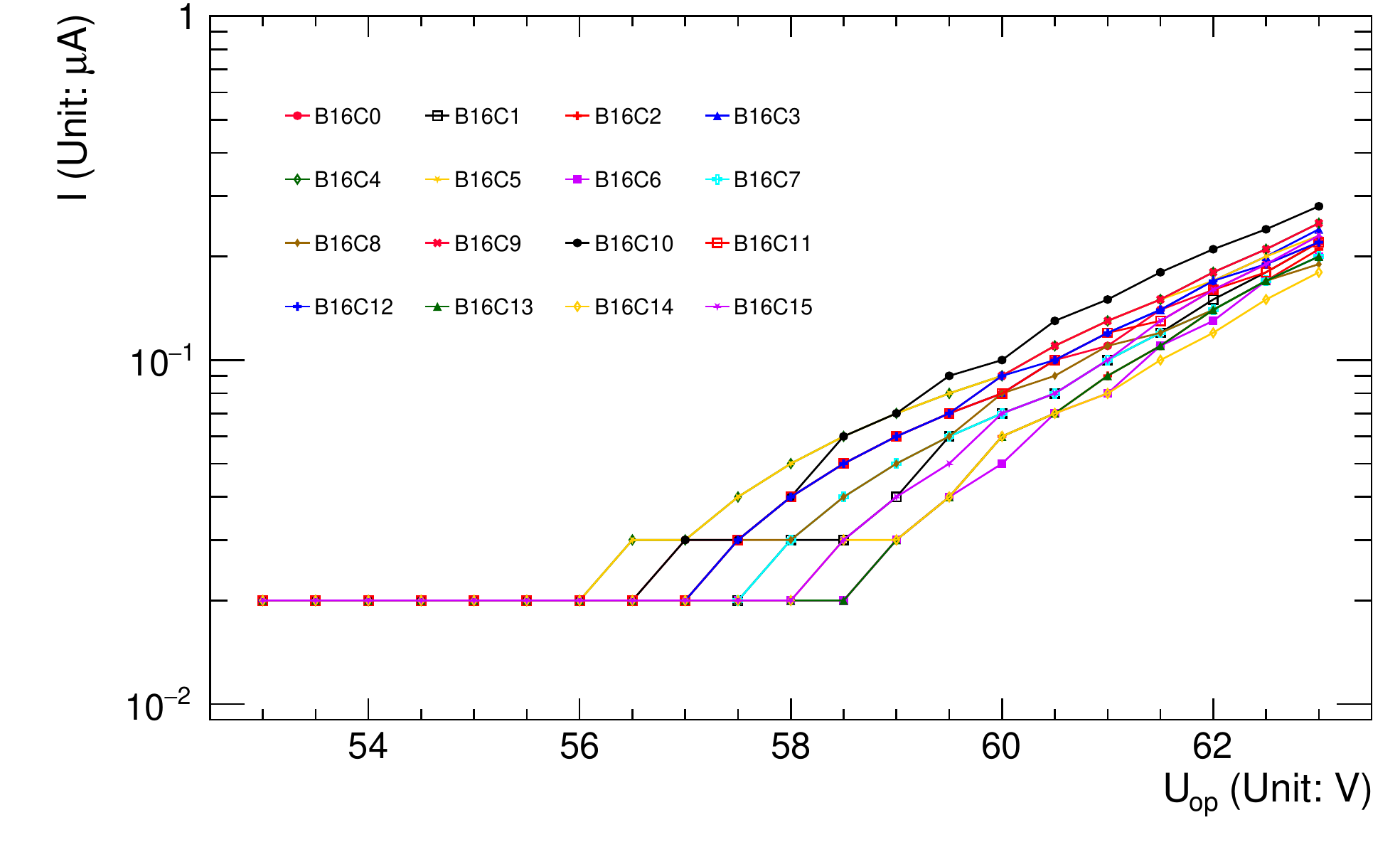}
    \caption{U-I curves of 16 SiPMs on one board}
    \label{fig:UI16}
  \end{figure}

\section{QA Batch Test and Result}
\label{sec:BatchTest}
To accomplish the QA task in time for approximately 1000 channels of SiPMs and the corresponding 60 (approx.) SiPM boards, FEE boards and RX boards, we established a dedicated batch test setup, by which we could test 16 channels quickly in one run.
Figure~\ref{fig:System_Chart} shows the flowchart of this system.
A waveform generator (LeCroy ArbStudio 1104) provided the trigger for the DAQ system and drove the light-emitting diode (LED) light source.
The peak emission wavelength of the LED was approximately 470-480 nm, which matched the maximum detection efficiency of the S13360-1325PE SiPM.
A highly transparent hexagonal quartz prism was used to diffuse the LED light injected from one end.
A bundle of 24 clear fibers receive the relatively uniform light output from the other end of the quartz prism, with the optical coupling via the silicon oil.
Sixteen out of the 24 fibers from the inner part of the bundle were chosen to illuminate the SiPMs under the test, with a thin air gap between them. 
The LED, the quartz diffuser and the fibers were placed in a light-tight blackened Aluminum box, whose photo and inner structure are shown in Figure~\ref{fig:Light_Source}. 
The bias HV for SiPM/FEE was provided by the CAEN module N1470A, which was set at  -90 V.
Two Keithley 2231A-30-3 DC power suppliers offered required low voltages (LV) for the FEE ($\rm \mp 6 V$) and RX ($\rm \mp 6.5 V$) boards. 
Despite their relatively low accuracy the CAEN N6742 digitizer was chosen over the oscilloscope to record the signal waveform.
The main reason was that the digitizer could measure 16 channels simultaneously, which was four times faster than the oscilloscope and important for our QA task in the limited time. 
Throughout the batch test, the measurements were performed at the laboratory for the photoelectric device study, with controlled temperature of approximately $\rm 25 ^\circ C$ and a relative humidity of approximately $\rm 50\%$.

All the QA tests described in the previous section were applied in a sequential mode.
The following steps were proformed:

(1) Before doing any other QA tests, we used a microscope to visually inspect the mechanical aspects of the SiPMs, and check whether their positions on the board fitted the spacer well. The height of SiPMs should not exceed the thickness of the spacer (approx. 2 mm).

{\color{black}(2) Insert the SiPM Board to the FEE board, set bias HV at 40 V to simply check communication of each part and functionality of the system.}

(3) We first set the bias voltage at 46 V to ensure no SiPM signal would be produced.
Under this voltage, the pedestal offsets and noise spectrum could be measured.
We recorded 3000 waveforms with the offset DAC values setting at -128, 0 and 127 respectively.
These waveforms showed how the offsets changed with the DAC settings.
Figure~\ref{fig:waveform_DAC} shows the baseline level of the waveform, which depicts a small but none-zero slope.
However this slope only affects the determination of the pedestal slightly, and is negligible because the range of the offsets is much larger.
The dependence on the DAC setting of all 16 channels of the FEE board under test is shown in Figure~\ref{fig:ped_DAC}.
Obviously, all the 16 channels show similar pedestal-DAC relations, all of which contain the pedestal 0, as required.
Also, by performing a fast Fourier transform (FFT) on the recorded waveforms, we can obtain the average noise frequency spectra, as shown in Figure~\ref{fig:Digitizer_Noise}.
Different DAC settings do not affect the spectra because they only change the baseline level.

(4) After measuring the noise spectrum, we scanned the bias voltage from 50.0 V to 63.0 V in steps of 0.5 V to obtain the U-I curves of SiPMs.
A set of typical U-I curves measured for 16 channels of one SiPM board are shown in Figure~\ref{fig:UI16}.
All the U-I curves exhibit consistent shapes, with only modest differences on $V_{br}$.
As discussed in section~\ref{sec:ElectronicsQA}, we take the bias voltage at $I = 0.03 \mu A $ as $V_{op}$, and require this voltage to be within the range of 50.0 V to 60.0 V.

(5) {\color{black} We finally set the SiPM bias voltage to 60 V to analyze the signal characteristics.
In fact, 60 V is not the nominal $V_{op}$ for the SiPMs used by EPD, but rather it is the maximum bias voltage permitted.
However running SiPMs under this voltage for a short period of time in the QA test should not be a problem, and a high gain at 60 V also eases the analysis of the test data. 
The charge of a signal is determined by integrating the waveform over time, with the baseline subtracted.
The effect of the slope of the baseline is reduced by fitting the baseline with a linear function.
Once the charge distributions are obtained, they are fitted by a multi-Gaussian function to resolve the position and resolution ($\sigma$) of the pedestal and multi-photoelectron peaks.
A gain greater than $ 5\times10^{5}$ and an SPE energy resolution less than $ 40\%$ are required.
A typical measurement of the signal waveform and the charge distribution are shown in Figure~\ref{fig:SPE}.}

\begin{figure}[ht]
  \begin{floatrow}
  \centering
  \ffigbox{\caption{GUI control panel}\label{fig:ctrl}}{\includegraphics[width=0.3\textwidth]{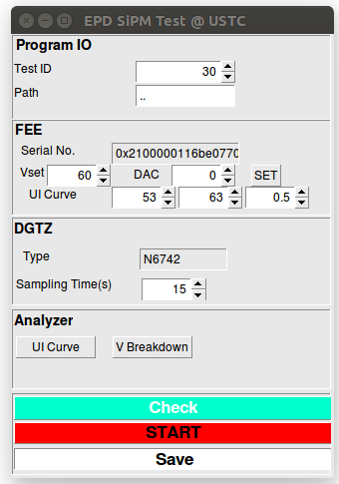}}
  \ffigbox{\caption{Database GUI}\label{fig:db}}{\includegraphics[width=0.49\textwidth]{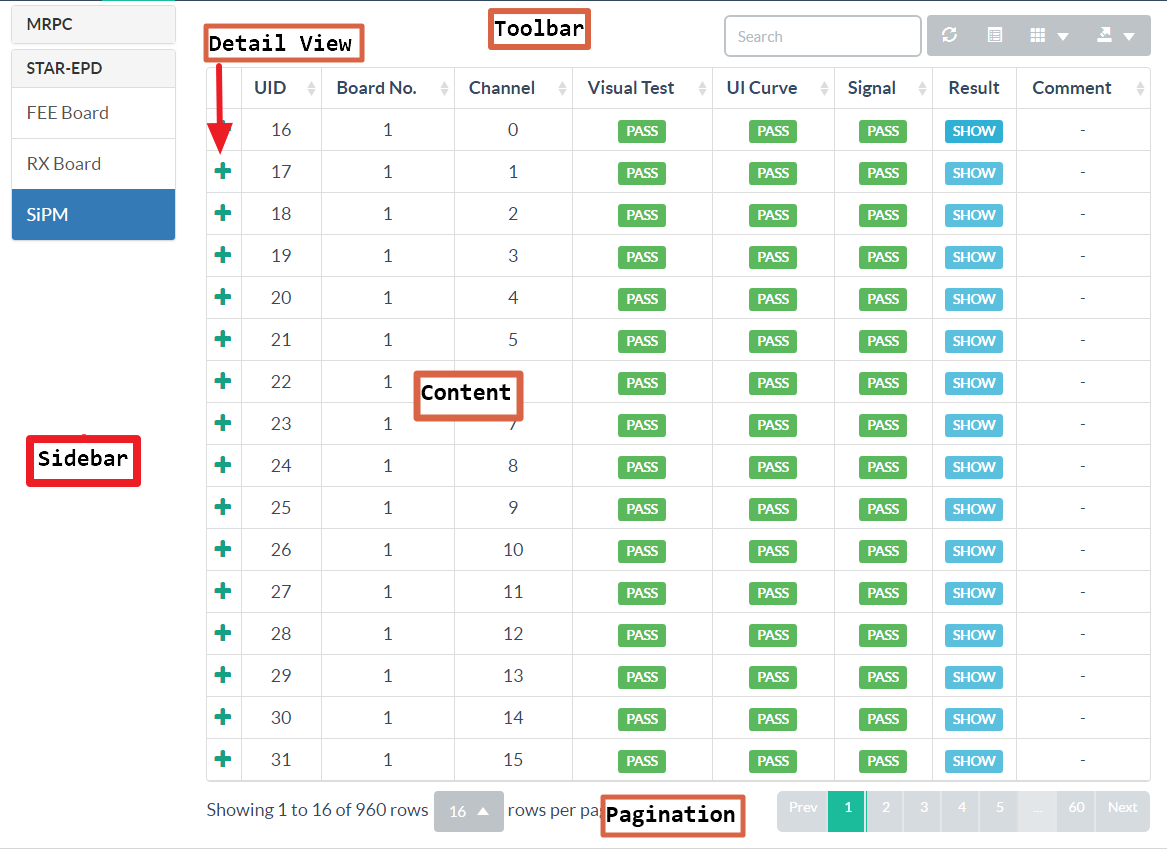}}
  \end{floatrow}
\end{figure}
To standardize and speedup the QA procedure, we developed a semi-automatic DAQ program using a GUI control panel.
The controller of this system is a C++ script program, which is based on the ROOT package.
It was developed to encapsulate the drivers of the FEE and the digitizer, and to provide a graphical interface for easy use.
As shown in Figure~\ref{fig:ctrl}, there are five sections in the GUI, which contains a user interface of the test information, hardware settings, and action buttons. 

A database was established to save the QA test data.
Based on the QA criteria, the qualities of the SiPMs and the FEE boards were examined, and the QA results were stored.
To organize the data in a manner that allows us to store, query, sort, and manipulate the data, we developed a simple database called DetectorDB, which was launched on our website.
The DetectorDB provides a web viewer to browse the records and test results in the database.
It consists of a main table (bootstrap-table), an image viewer (baguetteBox.js), and a ROOT file browser (jsroot).
An example of a page of the database is shown in Figure~\ref{fig:db}.
For more details about the implementation, one may visit the corresponding repository on GitHub\footnote{https://github.com/yatowoo/DetectorDB}.

\begin{figure}
    \centering
    \begin{floatrow}
        \centering
        \ffigbox{\caption{Operating Voltage by using the inflection point of UI curve; Blue: Data from Hamamatsu in which $V_{op}$ is defined as $V_{br}+5V$, Red: Data from U-I curve.}\label{fig:qc-vop}}{\includegraphics[width=0.45\textwidth]{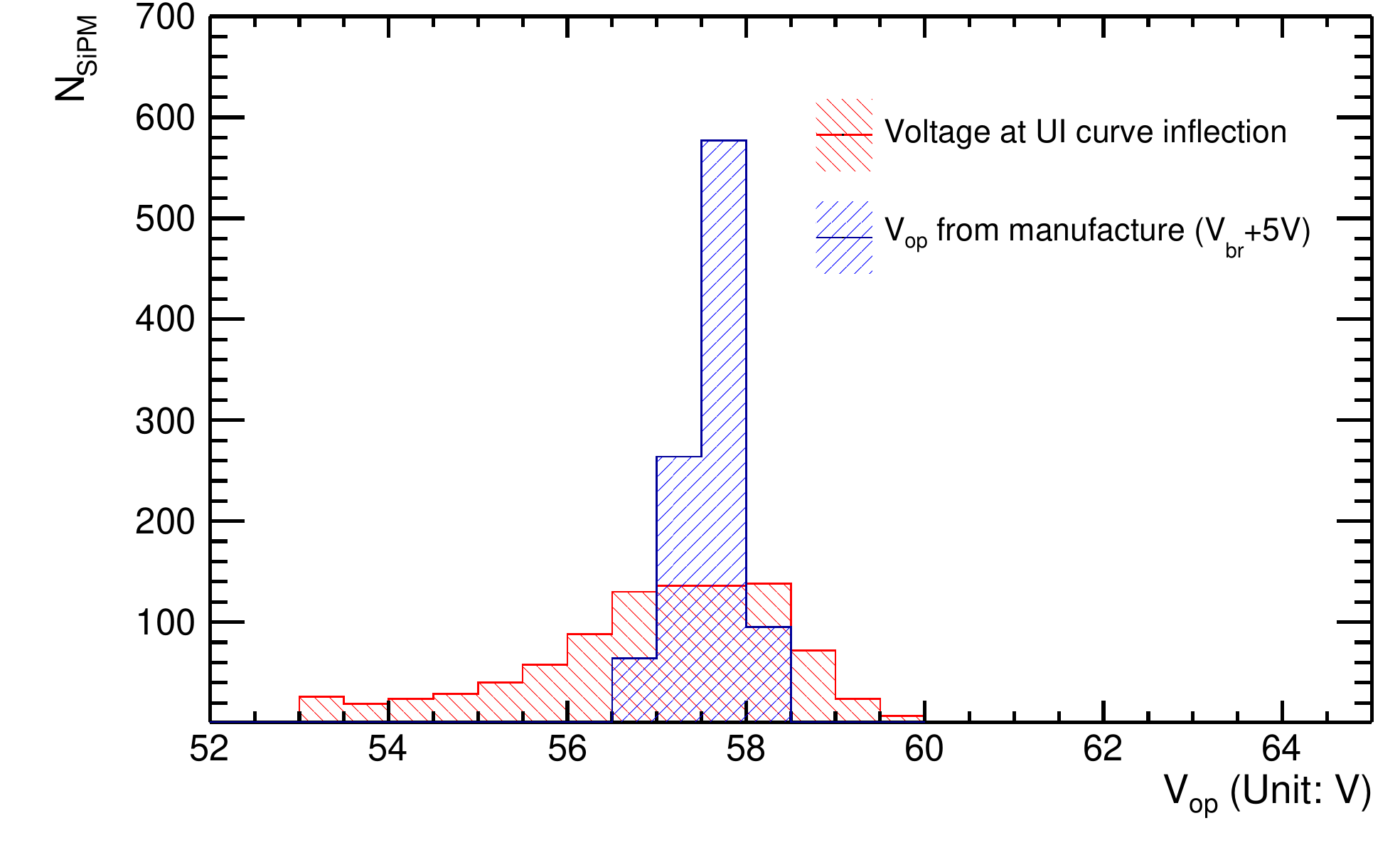}}
        \centering
        \ffigbox{\caption{Integrated plot for the charge spectra of all SiPMs with bias voltage at 60V.}\label{fig:qc-spec}}{\includegraphics[width=0.45\textwidth]{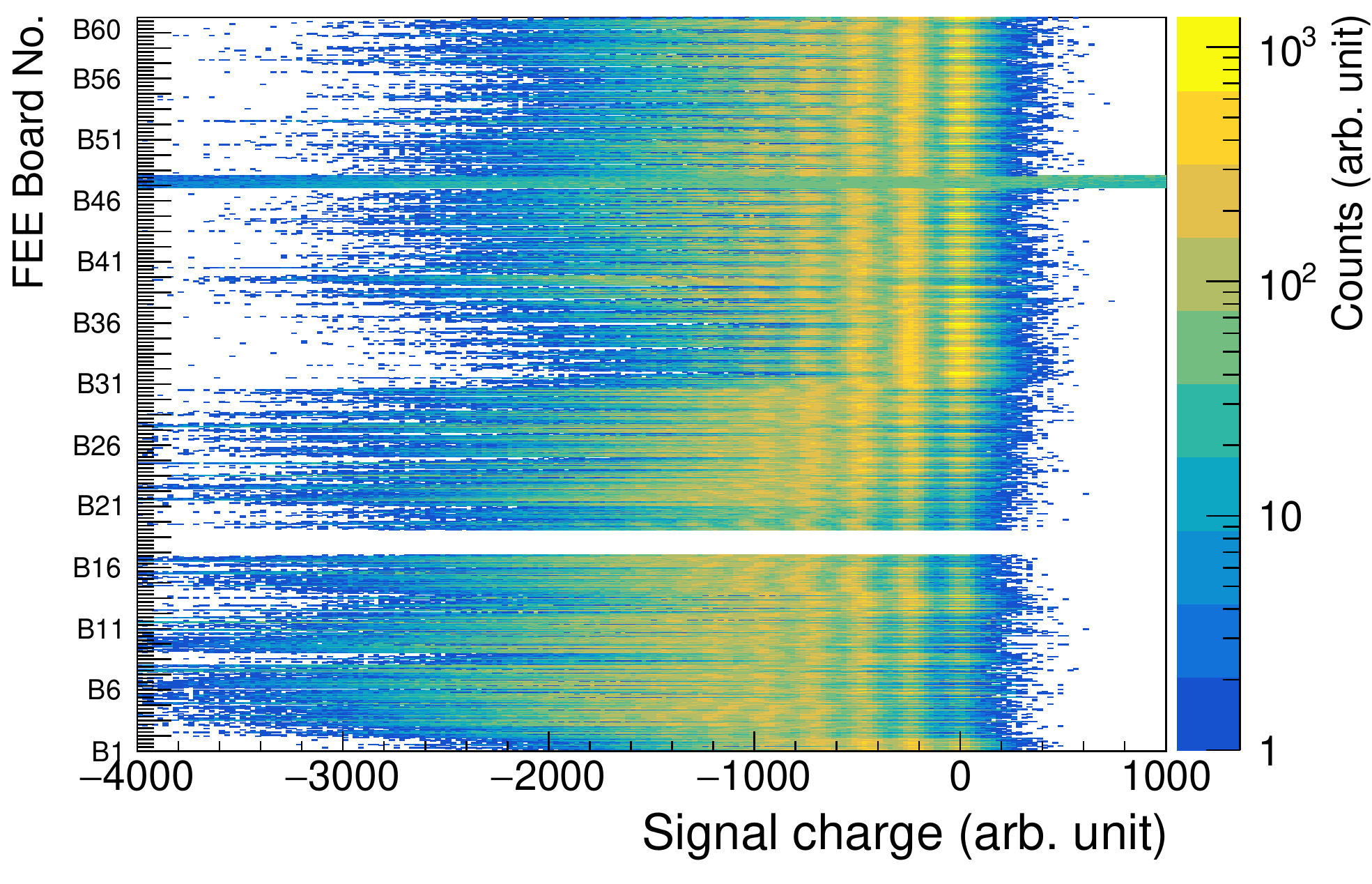}}
    \end{floatrow}
\end{figure}

\begin{figure}
  \centering
  \begin{floatrow}
    \centering
    \ffigbox{\caption{All SiPMs' gain measured by digitizer.}\label{fig:Gain-all}}{\includegraphics[width=0.45\textwidth]{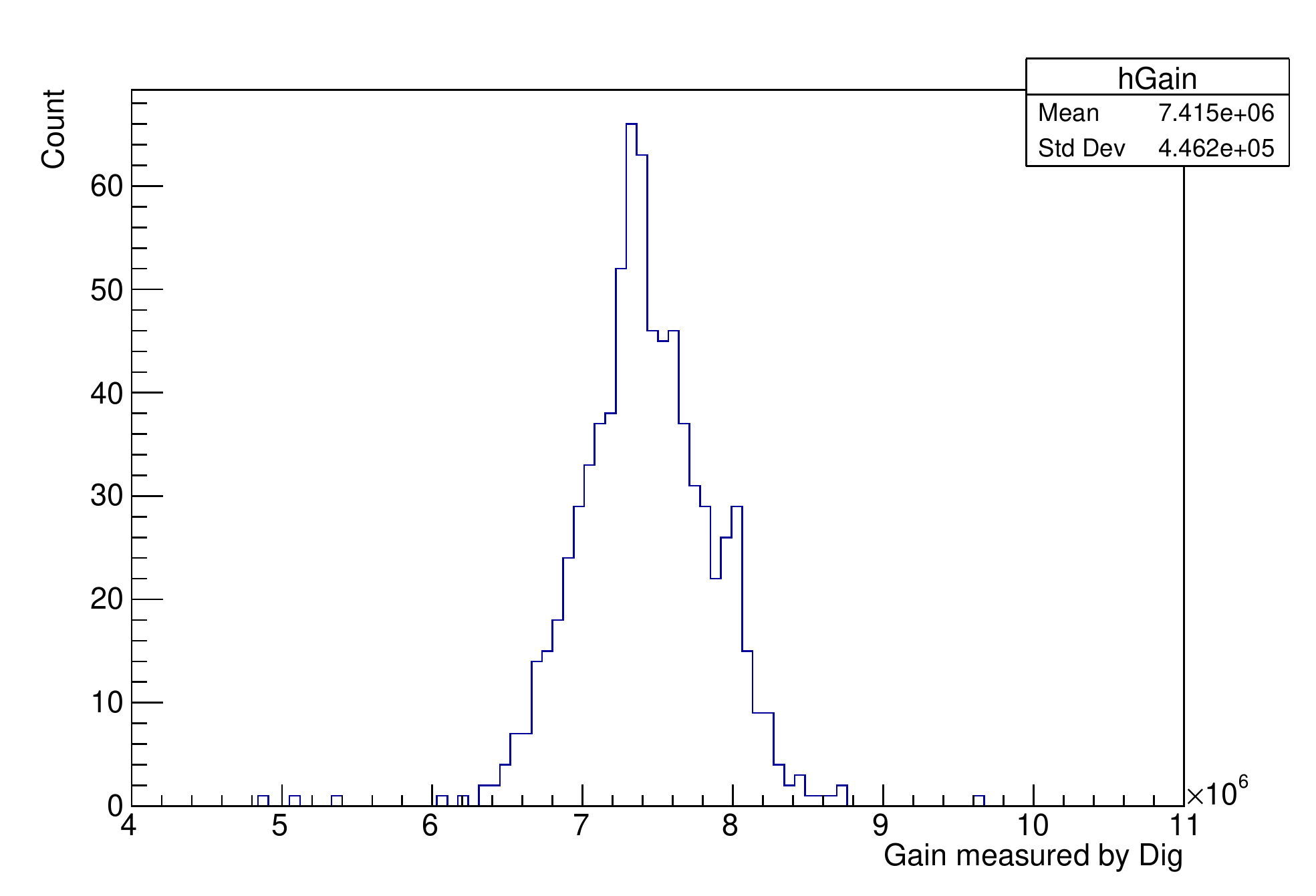}}
    \ffigbox{\caption{All SiPMs' SPE resolution $R_{SPE}$.}\label{fig:Res-all}}{\includegraphics[width=0.45\textwidth]{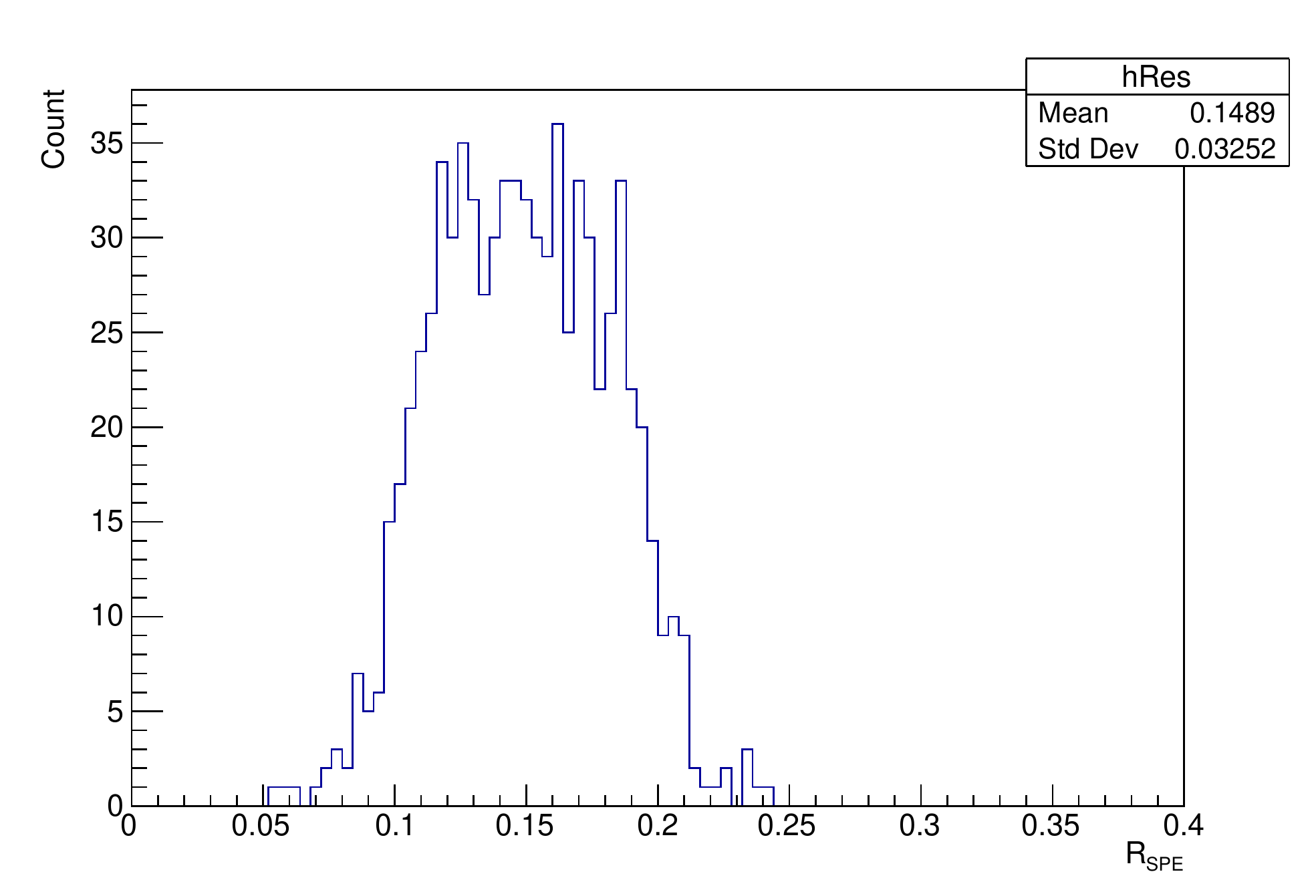}}
  \end{floatrow}
\end{figure}


\begin{table}
  \caption{QA Batch Test Items and Results}
  \label{table:QA Table}
  \begin{tabular}{|p{2.5cm}|p{4cm}|p{2.5cm}|p{5cm}|}
    \hline
    Items&Methods&Criteria&Result\\
    \hline
    SiPM assembly & Visual inspection with a microscope & Height $\in$ [0.83, 0.93]mm, soldered properly, no visible pollution & 1 SiPM (out of 960) was soldered badly. The SiPM board containing this SiPM was marked out. \\
    \hline
    FEE communication & Send/receive command/data via DAQ & Can read serial NO., set HV/offset & Only 1 FEE board was unable to connect to DAQ\footnote{Has been repaired after sending to BNL.}\\
    \hline
    RX board signal & Feed output signal to oscilloscope, measure noise spectrum & No unreasonable peak in the noise spectrum & Only 1 channel in one RX board was unable to output the signal\\
    \hline
    SiPM {\color{black}operating} voltage & Measure the U-I curve & $V_{op}\in (50V, 60V)$, {\color{black}$I_{leak}<0.03\mu A$} & All SiPMs pass the test\\
    \hline
    Single photon charge resolution & Collect waveforms by the digitizer and translate into the charge spectrum off-line & $R<0.4$ & All SiPMs and electronic devices passed the test \\
    \hline
    Signal offset & Collect waveforms by the digitizer and measure the offset off-line & Offset can be adjusted to 0 & All electronic devices passed the test\\
    \hline
    SiPM and FEE noise & Collect waveforms by the digitizer, do FFT off-line & No unreasonable peak in the FFT spectrum & All SiPMs and FEE boards pass test\\

    \hline
  \end{tabular}
\end{table}

All the QA items, criteria and statistical results are presented in Table ~\ref{table:QA Table}.
{\color{black}
On testing all the 60 SiPM boards and 960 SiPM channels, only one SiPM was not soldered properly, and the related SiPM module was rejected by the QA test. One FEE board was unable to connect to the DAQ, while one RX board contains a channel with no signal output.
Very consistent performance was shown by 57 out of the 60 electronic suite of SiPM-FEE-RX boards, and these boards passed the test.
Since STAR only requires 52 boards including 4 for 2 extra super-sectors, there are still enough good boards (57) to fulfill the standard of failure rate. However, the problematic boards have been fixed by the BNL electronics group after further investigations.
}

As a key parameter, we focused on the distribution of the operating voltages and compared them to the distribution of the breakdown voltages measured by the manufacturer (see Figure~\ref{fig:qc-vop}).
{\color{black}
Figure~\ref{fig:qc-spec} illustrates an integrated 2D histogram depicting the signals of all SiPMs at 60V, which is also the minimum value to cover the distribution of $V_{op}$ (see \ref{fig:qc-vop}).
Most channels have at least three peaks that can be easily distinguished by eye.
The two regimes with an overall shift, which are divided by the boundary between B30 and B31, should be caused by an adjustment of LED amplitude. It was proposed to reduce the luminous flux and to extract higher value at the first three peaks.
Beyond all that, we found some boards failing the test because of fault soldered joint (B17), connection error (B18) and noisy response (B47).}
{\color{black} Figure~\ref{fig:Gain-all} and Figure~\ref{fig:Res-all} are gains and $R_{SPE}$ results that calculated from charge spectra (see Section 3). Due to the excessive operation voltage, the SiPMs show the atypically high gain around $7\--8\times 10^{6}$ rather than the nominal  $0.5\--1\times 10^{6}$. Nevertheless, the distribution of single photon charge resolution displays an excellent performance at range from 0.05 to 0.25, which is much better than the criteria of 0.4.}
\section{Summary}
\label{sec:Conclusion}

The production and installation of the EPD sub-system into the STAR experiment is an important upgrade for the heavy-ion physics program at RHIC.
To meet the scheduled BES-II program, we need to efficiently manage and reliably excute the mass production of all EPD components, including the detectors, the readout electronics, and the integration and commissioning of the whole system.
The heavy-ion group at USTC has taken the task of producing all the FEE and the related QA testing to ensure the proper functionalities of the photon sensors (SiPMs) and the three electronic boards.
Based on the experience learned from a previous study of the EPD prototypes~\cite{EPD_proposal}, a dedicated QA procedure was established and applied to all SiPMs and electronic boards, up to $\sim 1000$ channels.
The QA test results revealed that our mass production and quality control action were very successful, with more than $98\%$ of these channels passing the QA requirements.
Very few channels faced any problems.

The full production of EPD was launched in the early of 2017 and completed by the end of that year.
From the 2018 to 2019 operations at RHIC, the EPD sub-system was continuously optimized and adjusted.
Most of the EPD components, including the FEEs, worked well and remained stable.
The results of the data analysis prove that the EPD performance fulfilled all the technical and physical requirements~\cite{EPD_NIM}. 
The EPD continues to run along with the STAR experiment.


\end{document}